\documentclass[draftclsnofoot,12pt,onecolumn]{IEEEtran}
 
\usepackage{amsmath}
\usepackage{amsthm}
\usepackage{amsfonts}
\usepackage{amssymb}
\usepackage{makeidx}
\usepackage{cite}
\usepackage{graphicx,bigints}
\usepackage{mathtools}
\usepackage{cases}
\usepackage{color}
\usepackage{hyperref}
\usepackage{bbm}
\usepackage[normalem]{ulem}
\usepackage{braket}
\usepackage{kotex}
\usepackage{epstopdf}

\newtheorem{theorem}{Theorem}

\newtheorem{lemma}{Lemma}

\newtheorem{example}{Example}

\newtheorem{proposition}{Proposition}
\newtheorem{remark}{Remark}

\DeclareMathOperator{\cV}{\mathcal{V}}

\DeclareMathOperator{\cD}{\mathcal{D}}
\DeclareMathOperator{\cT}{\mathcal{T}}

\DeclareMathOperator{\SIR}{\textrm{SIR}}
\DeclareMathOperator{\bR}{\mathbb{R}}

\DeclareMathOperator{\bP}{\mathbf{P}}
\DeclareMathOperator{\ind}{\mathbbm{1}}
\DeclareMathOperator{\bE}{\mathbf{E}}
\DeclareMathOperator{\bZ}{\mathbb{Z}}
\newcommand*\diff{\mathop{}\!\mathrm{d}}

\newcommand*\nnb{\nonumber}

\newcommand\independent{\protect\mathpalette{\protect\independenT}{\perp}}
\def\independenT#1#2{\mathrel{\rlap{$#1#2$}\mkern2mu{#1#2}}}

\newcommand{\ea}{\stackrel{(\text{a})}{=}}
\newcommand{\eb}{\stackrel{(\text{b})}{=}}
\newcommand{\ec}{\stackrel{(\text{c})}{=}}

\title{Analysis of Vehicular Safety Messaging in Cellular Networks}

\author{Chang-Sik~Choi and Fran{\c{c}}ois~Baccelli
\IEEEcompsocitemizethanks{\IEEEcompsocthanksitem 	{Chang-sik Choi is with Qualcomm Inc., Bridgewater, NJ 08807. Francois Baccelli is with Wireless Networking Communication Group, Department of Electrical and Computer Engineering and Department of Mathematics, The University of Texas at Austin, Austin, TX 78712 (email: chang-sik.choi@utexas.edu, baccelli@math.utexas.edu)} }
\IEEEcompsocitemizethanks{\IEEEcompsocthanksitem}
}
\begin{document}
	\maketitle
\begin{abstract}	
    This paper concerns the performance of vehicle-to-everything (V2X) communications. More precisely, we analyze the broadcast of safety-related V2X communications in cellular networks where base stations and vehicles are assumed to share the same spectrum and vehicles broadcast their safety messages to neighboring users.  We model the locations of vehicles as a Poisson line Cox point process and the locations of users as a planar Poisson point process. We assume that users are associated with their closest base stations when there is no vehicle within a certain distance $ \rho $. On the other hand, users located within a distance $ \rho $ from vehicles are associated with the vehicles to receive their safety messages. We quantify the properties of this vehicle-prioritized association using the stochastic geometry framework. We derive the fractions of users that receive safety messages from vehicles. Then, we obtain the expression for the signal-to-interference ratio of the typical user evaluated on each association type. To address the impact of vehicular broadcast on the cellular network, the paper also derives the effective rate offered to the typical user in this setting. 
\end{abstract}
\section{Introduction}
\subsection{Motivation and Related Work}
Device-to-device (D2D) technology has been used in various applications including file transfer \cite{6953022}, emergency service \cite{6933414}, and intelligent transportation systems (ITS)\cite{7017566}. In the context of cellular 5G NR vehicle-to-all (V2X) applications \cite{38885}, vehicle or pedestrian user equipment (UE) is enabled to transmit and receive data to/from nearby vehicular or roadside unit (RSU) UE using direct sidelink communications in the presence of existing cellular uplink or downlink \cite{6515060,7992934}. The messages transmitted through sidelink communications include basic safety information as well as traffic information \cite{kenney2011dedicated}. In safety applications,  it is essential for other vehicular or pedestrians UEs within a certain distance to be able to decode the data transmitted through sidelink even when this type of communications coexist with the existing cellular infrastructure \cite{7017566,6515060,7992934,36885,38885}. This paper analyzes the performance of sidelinks and cellular communications when sidelinks are prioritized over cellular for users close to vehicles.

\par The representation of the locations of D2D transceivers based on planar Poisson point processes has been widely used in the literature  (See \cite{baccelli2010stochastic,6042301,6524460,6566864,6515339} and references therein). The typical user performance was analytically derived in \cite{baccelli2010stochasticvol2}. Moreover, various network protocols and techniques such as power control \cite{6133537,6928445}, multi-input-multi-output \cite{6047553,7042756}, and access control \cite{7056528} were investigated in this context. The planar Poisson point process is tractable because its points are spatially independent \cite{daley2007introduction,chiu2013stochastic}. In networks with vehicles, on the other hand, vehicles exhibit a linear co-location and  their motions are constrained by road layouts.  The Poisson point process, however, is unable to capture such spatial constraints \cite{chiu2013stochastic}. As an alternative, Poisson-line Cox point processes were used to model vehicles on roads \cite{baccelli1997stochastic,morlot2012population,8357962,8419219,8340239,8796442}. In  Cox-based vehicle models, road networks are first modeled by a stationary Poisson line process and, conditionally on lines, vehicles are modeled by Poisson point processes on the lines. These models inherently capture the relationship between vehicles and their corresponding road structure.

\par In particular, the Cox-based model is useful to the analysis of V2X networks when vehicular networks coexist with cellular networks \cite{36885,38885}. In \cite{morlot2012population,8357962,chetlur2019coverage,8796442}, the V2X network was represented by two distinct spatial architectures; the cellular network is modeled by a planar Poisson point process and the vehicle network is modeled by an independent Poisson-line Cox point process .  The distributions of signal-to-interference-plus-noise ratio (SINR) of typical links were analyzed in these papers. Particularly, \cite{8357962}  derived the SINR distributions of all existing typical pairs in such networks: links from vehicle base stations to vehicle users, from vehicle base stations to planar users, from planar base stations to planar users, and from planar base stations to vehicle users. In \cite{8357962}, users are assumed to be associated with their nearest transmitters to increase the data rate or coverage area \cite{6171996,6524460}. 

\par In safety applications, each vehicle {broadcasts} its basic safety information such as its location, speed, and acceleration to nearby pedestrians or vehicles. Specifically, if pedestrians or cellular user are within a certain distance from a vehicular UE, they are required to decode the messages from vehicles \cite{7017566,36885,38885,6515060,7992934,kenney2011dedicated,vanetWD12}. For instance, when ITS with vehicular safety messaging is deployed with a cellular architecture, the users are prioritized to be associated with vehicles over cellular base stations as long as they are close to vehicles. The aim of this paper is to evaluate the reliability of the communications from vehicles to users as well as the effective rate of downlink from cellular base stations to users. To the best of our knowledge, the interplay between sidelink broadcast and downlink unicast has not been investigated within the stochastic geometry framework. 




\subsection{Contributions}
\textbf{Modeling of vehicular broadcast in cellular networks}: We model the locations of base stations as a planar Poisson point process of intensity $ \lambda_b $, and the location of cellular downlink users as an independent planar Poisson point process of intensity $ \lambda_u $. To represent the linear patterns of vehicles on roads, we use an isotropic Poisson line process to model the road layout and then use Poisson point processes  conditionally on the line process to model the locations of vehicles. We assume that control messages from vehicles have users decode vehicular safety messages when the users are within a distance $ \rho $ from vehicles. 
Fig. \ref{fig:combineddeployment} illustrates such vehicle-prioritized association region.

\textbf{Derivation of network performance from a typical user perspective}: The stationary framework of the proposed model allows us to derive the network performance of a typical user, which corresponds to the network performance spatially averaged over all users in the network. Specifically, we consider a V2X system where the typical user association is determined by the locations of vehicles and base stations with respect to the typical user. Using the mean area fraction of the Boolean model---which quantifies the relative size of the vehicular broadcast region in the Euclidean plane---we derive the probability that a typical user at the origin is prioritized to receive the vehicular broadcast message or is associated with a cellular base station for downlink communications, respectively. Conditional on the association of the typical user, we derive its signal-to-interference (SIR) distribution in terms of an integral formula. The impact of vehicular broadcasts on cellular unicasts is characterized as the effective data rate of cellular links, which accounts for the long-term average rate that the cellular users are offered through downlink unicast, assuming that each downlink base station is equally shared by all of its associated users. For this, we derive a formula for the mean number of users associated with each base station, by quantifying the area of the vehicle association region inside a Voronoi cell of the base station point process.

\section{System Model}
This section discusses the spatial model, the vehicle-prioritized association principle, the propagation model, and the network performance metrics.
\subsection{Spatial Model}\label{S:2-A}
We model the random locations of cellular users, namely UEs in cellular networks, as a planar Poisson point process $ \Phi_u $ with intensity $ \lambda_u $. The base stations of the cellular networks are modeled by an independent planar Poisson point process $ \Phi_b $ with intensity $ \lambda_b. $  The locations of users and base stations are fixed (do not change overtime). In addition, we assume the density of users is high, i.e., $ \lambda_u>\lambda_b, $ and also assume that there is at least one active downlink user in each base station cell.

\par We model the locations of vehicles as a Cox point process $ \Phi_v$ \cite{8419219} based on a Poisson line process. Roads are modeled by a stationary and isotropic Poisson line process $ \Phi_l $. This line process is given by a Poisson point process on the cylinder $ \mathbf{C}:=\bR\times (0,\pi) $ with intensity $ \lambda_l/\pi$ \cite{chiu2013stochastic}. Specifically, a point of the cylinder Poisson point process, say $ (r,\theta)\in\mathbf{C}, $ produces an undirected line on the Euclidean plane $ \bR^2  $. Here, $ r $ corresponds to the algebraic displacement from the origin to the line and $ \theta $ corresponds to the angle between the line and the positive $ x$-axis in the counterclockwise direction. Conditionally on the line point process, vehicles are modeled by an independent one-dimensional Poisson point process, say $ \phi(r,\theta), $ with intensity $ \mu $ on each line.  Consequently, the proposed vehicle point process is stationary  and the spatial density of vehicles per unit area is $ {\lambda_l\mu}/{\pi} $\cite{8419219}. As shown in the literature including \cite{8419219}, the Cox-based vehicle point process captures the relationship between vehicles and corresponding road layouts. 

\par This paper assumes that vehicles move at a constant speed $ v $ along the lines on which they are located.  For simplicity, we assume that the vehicles' moving directions are independently chosen with probability half. Once vehicles' moving directions are chosen at time $ t=0 $, they are maintained forever. In the remainder of this paper, the vehicle point process at time $ t $ is denoted by $ \Phi_{v;t}. $ 

\subsection{Channel Model}\label{Sec:2-B}
Let $ d $ denote the distance between a transmitter and its designated receiver, or receivers (for broadcast). In this paper, we assume a rich scattering around cellular users such that the received signal power is characterized by the product of a distance-based path loss and a small-scale Rayleigh fading \cite{goldsmith2005wireless,tse2005fundamentals}. 
\par Time is assumed to be slotted. For each time slot, the received signal power is given by $ pHd^{-\alpha} $ where $ p $ is the transmit power of the transmitter, $ H $ is an exponential random variable with mean one, and $ \alpha $ is the path loss exponent assumed to be greater than two. The transmit powers of base stations and vehicles are denoted by $ p_b $ and $ p_v, $ respectively. 
\begin{figure}
	\centering
	\includegraphics[width=.8\linewidth]{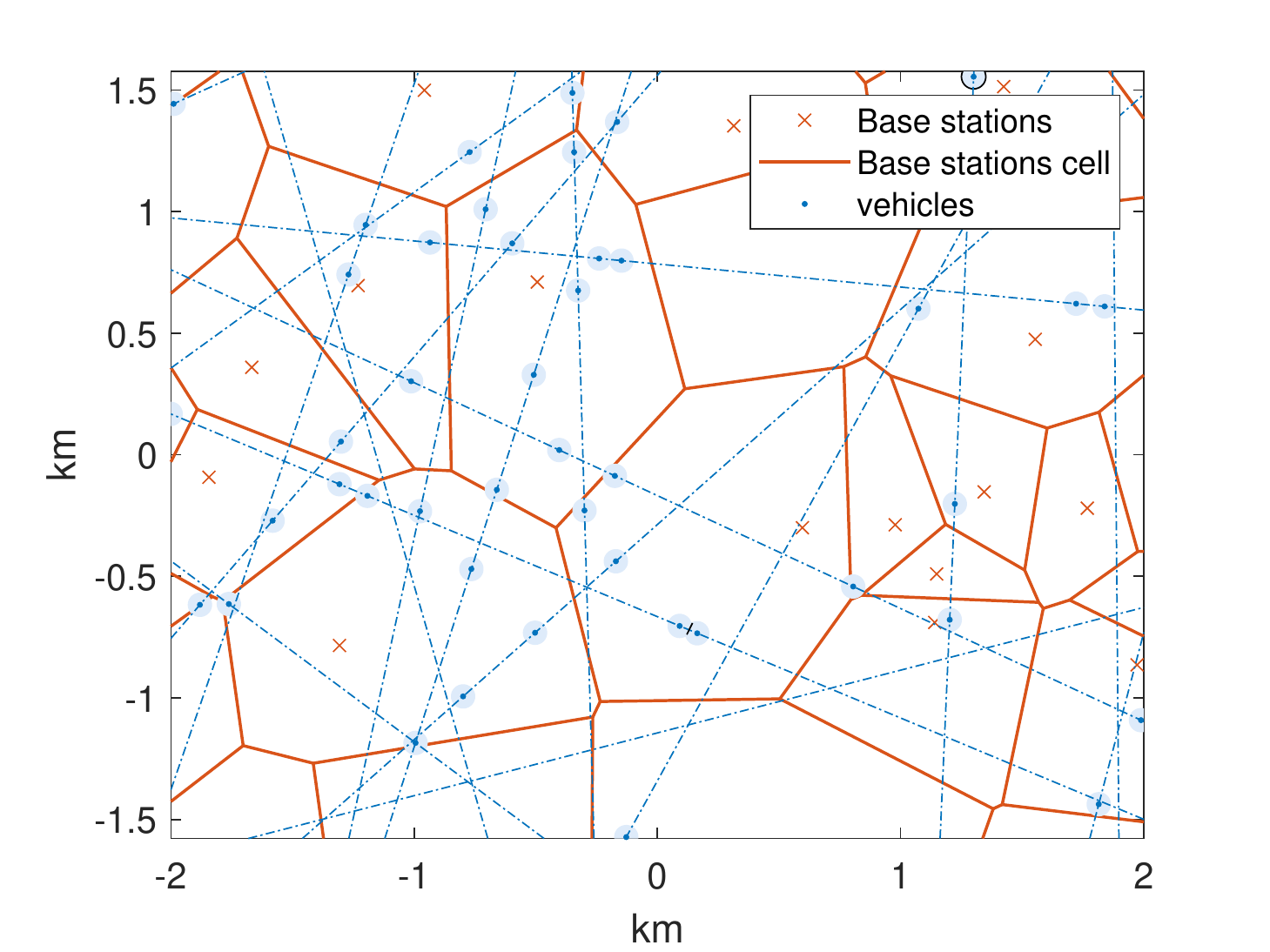}
	\caption{The vehicle association region is given by Boolean model in blue and the rest of the plane is subdivided in base station cells.  Users in the vehicle association region are configured to decode safety messages from vehicles. We consider  $ \lambda_l=5/\text{km}, $ $ \mu=1/\text{km} $, $ \lambda_b=1/\text{km}^2, $ and $ \rho=50$ meters.}
	\label{fig:combineddeployment}
\end{figure}

\begin{figure}
	\centering
	\includegraphics[width=0.8\linewidth]{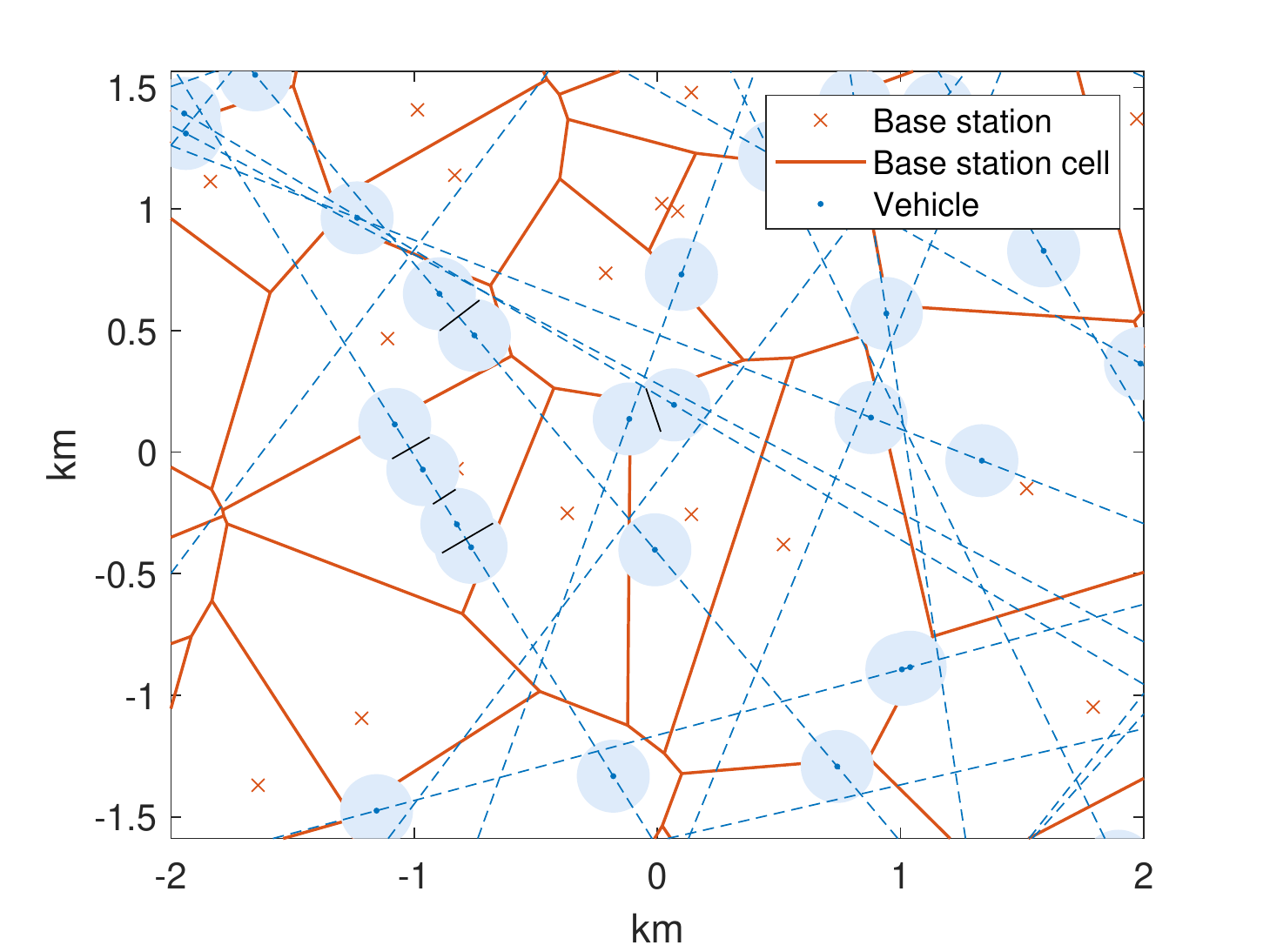}
	\caption{Illustration of vehicle association region and base station cells. We consider $ \lambda_l=4/\text{km}, $ $ \mu=0.5/\text{km} $, $ \lambda_b=1/\text{km}^2, $ and $ \rho=150$ meters. Solid black lines indicate the boundary of vehicular association region.}
	\label{fig:deployment2}
\end{figure}


\subsection{Vehicle Priotizied Association Principle}\label{SS:Association}
Note that in ITS, it is essential for cellular users to be able to decode sidelink messages if there is a vehicle within some distance \cite{kenney2011dedicated}. Thus, in this paper, we define the user association as follows: each user is associated with its nearest base station. However, if a cellular user is within some distance $ \rho $ from a vehicle, that user is associated with the corresponding vehicle. If there are two such vehicles, the user is associated with the closest one. 
\begin{remark}
	In practice, vehicle-prioritized association can be implemented using sidelink control messages. For instance, vehicles broadcast messages are preceded by vehicle control messages. Any user who receives a sidelink control message from a vehicle is required to try to decode the safety messages from the corresponding vehicle. In this context, the radius of the disk $  \rho $ specifies the maximum distance to the users who try to decode the safety message.  
\end{remark}

Figs. \ref{fig:combineddeployment} and \ref{fig:deployment2} represent the association regions in the Euclidean plane. The vehicle association region is illustrated as the union of disks centered at vehicles. Users within this region are associated with their corresponding vehicles. The vehicle association region is given by 
\begin{equation}\label{setD}
\cD(t)=\bigcup_{X_i\in\Phi_{v;t}}B_{X_i}(\rho),
\end{equation}
where $ B_x(r) $ denotes the disk of radius $ r $ centered at $ x\in\bR^2 $. On the other hand, the cellular association region is given by \begin{equation}\label{eq3}
\bR^2\setminus \cD(t).
\end{equation}


\begin{remark}
	In this paper, downlink and sidelink transmissions are assumed to differ in the way they deliver their messages. Specifically, downlink transmissions from base stations to cellular users are assumed to be \emph{unicasts}; each base station selects one of its associated users at each time slot and serves it. On the other hand, sidelink transmissions from vehicles to users are assumed to be \emph{broadcasts}; in other words, each vehicle broadcasts its message to nearby users and all of its users attempt to decode the message. 
\end{remark}




%
\begin{center}
	\begin{table}\caption{Spatial Model and Description}\label{Table:1}
		\centering
		\begin{tabular}{|l|l|}
			\hline
			Notation and distribution & Description\\ 
			\hline 
			$ \Psi \sim$ Poisson$ (\lambda_u )$ & User point process\\
			\hline 
			$ \Phi_b\sim $ Poisson$ (\lambda_b) $ & Cellular base stations\\ 
			\hline 
			$ \Phi_{v;t}\sim \text{Poisson line Cox} $& Vehicles  \\ 
			\hline 
			$\cD(t) \sim $ Boolean on $ \Phi_{v;t} $& Area assoc. w/ vehicles \\ 
			\hline 
			$ \bR^2\setminus \cD(t) $   & Area assoc. w/ base stations \\
			\hline			
			$\Psi\cap \cD(t)   $ &  Users  assoc. w/ vehicles \\ 
			\hline 
			$\Psi\cap (\bR^2\setminus \cD(t)) $&  Users assoc. w/ base station\\ 
			\hline 
			$\Psi_{v;t} \sim \text{Unif. on disks of } \rho \text{ at } \Phi_{v;t}  $ &  Users  served by vehicles at time t \\ 
			\hline
			$\Psi_{b;t} \sim $ Unif. on $ \cV\setminus \cD(t)$ at $ \Phi_b $  &  Users served by base stations at time t\\ 
			\hline 
		\end{tabular}
	\end{table}
\end{center}

\subsection{Preliminary: Typical Cell and Zero Cell}
The Voronoi tessellation $ \cV $ w.r.t. $ \Phi $ is the sequence of random sets as follows: 
\begin{align}
\cV=\{\cV_{i}\}_{i\in\bZ}=\{x\in\bR^2 s.t.   \|x-X_i\|<\|x-X_j\|, \forall X_j \in \Phi \setminus X_i\}_{i\in\bZ}
\end{align}
The points $ \{X_i\}_{i\in\bZ}$ are the nuclei of the Voronoi cells. 
Based on the association principle in Section \ref{SS:Association}, the cellular association region is the Voronoi tessellation w.r.t. the base station point process \emph{minus} vehicle association region. This is described as non-shared areas in figs. \ref{fig:combineddeployment} and \ref{fig:deployment2}.

\par Let us  consider the Palm distribution of the base station point process $ \Phi $ \cite{baccelli2013elements}, which has a typical point at the origin. Under this Palm distribution, the cell containing the typical point at the origin is referred to as a typical cell \cite{moller2012lectures}. 
\begin{figure}
	\centering
	\includegraphics[width=0.6\linewidth]{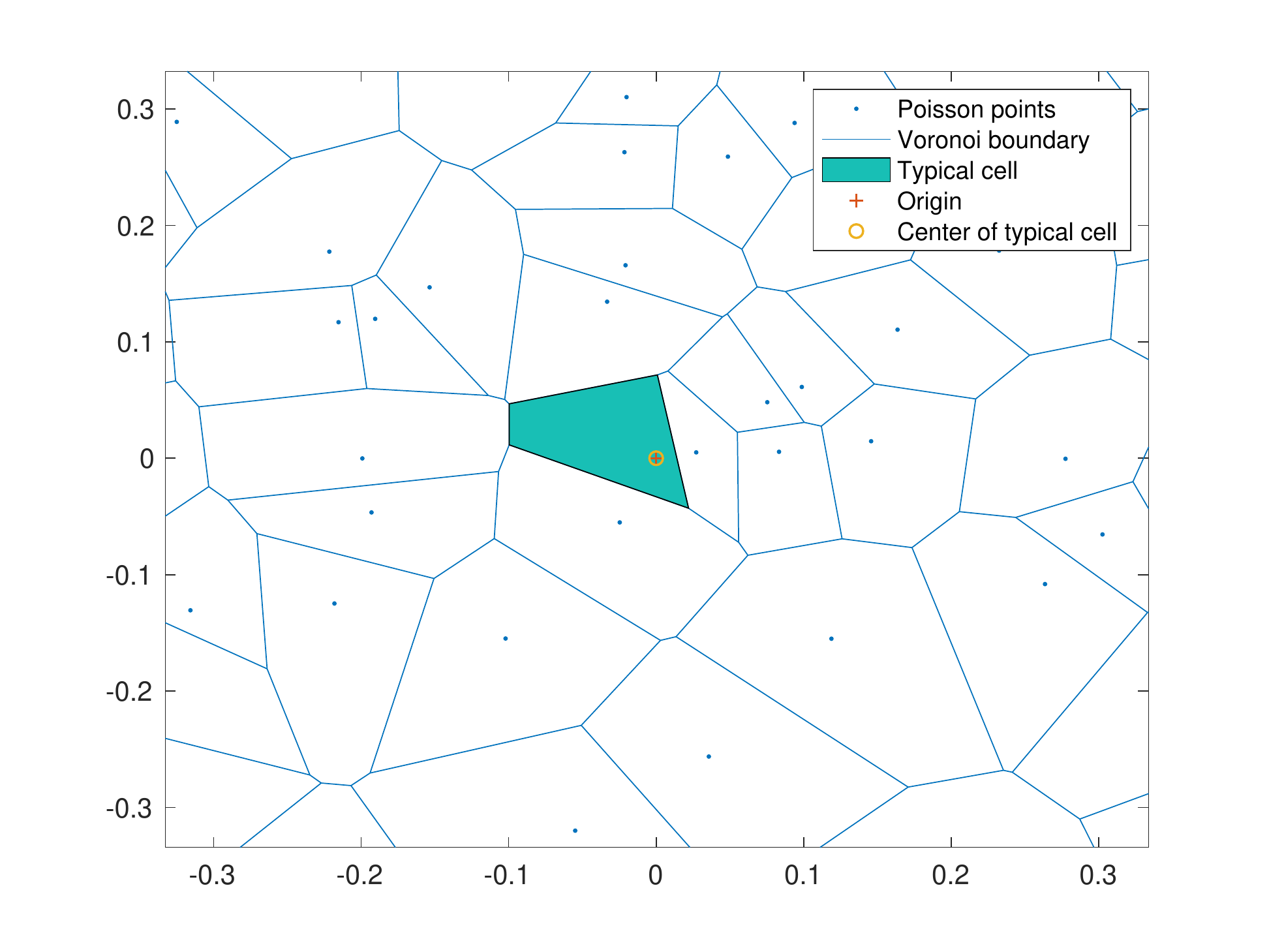}
	\includegraphics[width=0.6\linewidth]{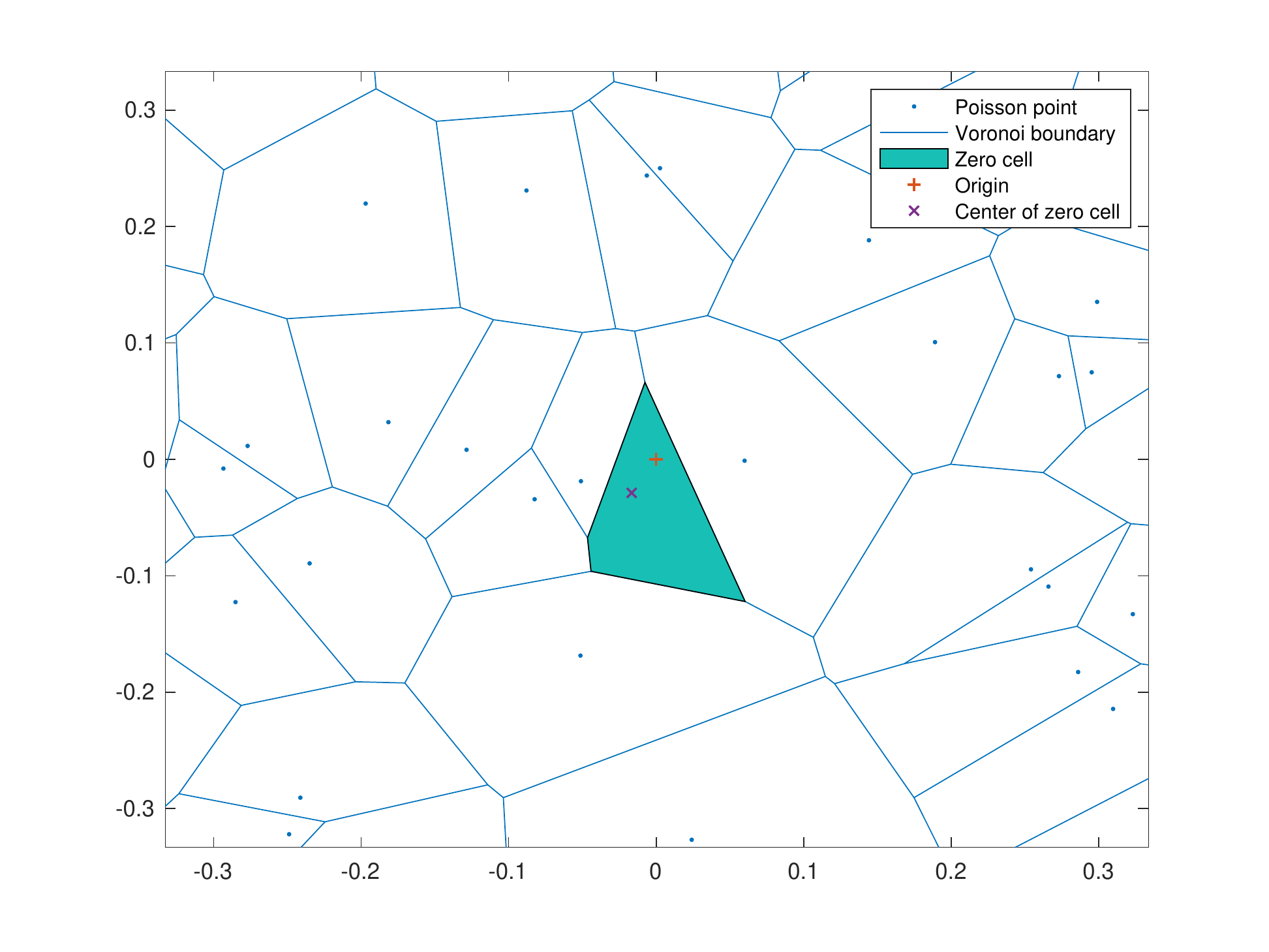}

	\caption{Illustration of the typical cell (upper) and the zero cell (lower). The typical cell center is the origin; the zero cell center is not the origin. The intensity of the Poisson point process is given as $ 50 $, for illustration.}
	\label{fig:zerocell}
\end{figure}
\par One may similarly consider the Voronoi tessellation of the base station point process $ \Phi $ under the stationary distribution of the user Poisson point process. Note that the origin is not an element of the base station point process. Instead, there exists a cell $ \cV_j $ that contains the origin (i.e., the typical user) and this cell is referred to as the zero cell. Fig. \ref{fig:zerocell} illustrates the typical cell (upper figure) and the zero cell (lower figure) w.r.t. the base station Poisson point processes of intensity $ 50/\text{km}^2  $, respectively. The zero cell will be used to derive the effective rate and then characterize the interplay between the cellular downlink unicast and vehicular sidelink broadcast. 

\subsection{Network Performance Metrics}\label{S:metric}
We first focus on the reliability of sidelink and downlink cellular communications, respectively. For each, reliability is captured by the coverage probability of the typical user. Then, we derive the effective data rate of the typical user, offered by downlink communications. 
\subsubsection{Coverage Probability}
We first use the Palm distribution of the user point process to feature a typical user at the origin. Note that the network performance seen by the typical user coincides with the network performance averaged over all users in the network \cite{baccelli2013elements}.  

\par  The coverage probability of the typical cellular downlink user $ \bP_{\Psi}^0(\SIR>\tau,\text{DL}) $ is defined by 
\begin{align}
\bP_{\Psi}^0\left(\left.\frac{p_bH\|X_0\|^{-\alpha}}{\sum_{X_i\in\Phi_{v;t}}p_vH{\|X_i\|}^{-\alpha}+\sum_{Y_i\in\Phi_{b}\setminus X_0}p_bH{\|Y_i\|}^{-\alpha}}>\tau\right.,0\notin\cD(t)\right),\label{eq:cov-def1}
\end{align}
where $ \tau $ is the decoding threshold, $ X_\star $ is the location of the base station that is closest to the origin, and $ \bP_{\Psi}^0 $ indicates that the metric is derived under the Palm distribution of $ \Psi $, namely for the typical user at the origin. In the proposed network, we assume that interference dominates thermal noise. 
Thus, we investigate the SIR coverage probability. 

\par Similarly, the coverage probability of the typical sidelink user $ \bP_{\Psi}^0(\SIR>\tau,\text{SL}) $ is defined by 
\begin{align}
	\bP_{\Psi}^0\left(\left.\frac{p_vH\|X_\star\|^{-\alpha}}{\sum_{X_i\in\Phi_{v;t}\setminus X_\star}p_vH{\|X_i\|}^{-\alpha}+\sum_{Y_i\in\Phi_{b}}p_bH{\|Y_i\|}^{-\alpha}}>\tau\right.,0\in\cD(t)\right),\label{eq:cov-def2}
\end{align}
where $ X_\star $ is the location of the nearest vehicle. 

\par By combining \eqref{eq:cov-def1} and \eqref{eq:cov-def2}, the coverage probability of the typical user is 
\begin{equation}
	\bP_{\Psi}^0(\SIR>\tau)=\bP_{\Psi}^0(\SIR>\tau,\text{DL})+\bP_{\Psi}^0(\SIR>\tau,\text{SL}).\label{eq:tpro}
\end{equation}

\subsubsection{Effective Data Rate}
The (load-balanced) effective downlink data rate is defined as the spatial average of the long-term achievable rate (bits/sec/Hz) of each cellular user, provided that the time resource of each base station is equally shared by all of its associated downlink users\footnote{A simple example: at each time slot, each base station chooses and serves one user out of its associated users.}. 

\par Conditionally on the downlink association, the mean achievable rate of the typical user is denoted by  $ \bE_{\Psi}^0[\log_2(1+\SIR)|\text{DL}] $. Here, we shall use the effective rate per user, which is defined as the mean achievable rate of the typical user, divided by the mean number of users that share the same base station. In addition, from the typical user's perspective, this effective rate is multiplied by the probability that that the typical user has a downlink association. Therefore, the effective data rate is defined as follows:
\begin{align}
{\cT}&=\bP(\text{DL})\cdot\frac{\bE_{\Psi}^0[\log_2(1+\SIR)|\text{DL}]}{\bE_{\Psi}^0[\|\Psi(\cV_{Z}\setminus \cD(t))\|]}=\frac{\bE_{\Psi}^0[\log_2(1+\SIR),\text{DL}]}{\bE_{\Psi}^0[\|\Psi(\cV_{Z}\setminus \cD(t))\|]}\label{eq:Throughput},
\end{align}
where $ \bE_{\Psi}^0[\|\Psi(\cV_{Z}\setminus \cD(t))\|] $ denotes the expected number of users that share the base station $ {Z} $ and  $ \cV_{Z}\setminus \cD(t) $ denotes the Voronoi cell of base station $ {Z} $ minus the vehicle associaiton region $ \cD(t)$. 
\begin{center}
	\begin{table}\caption{Performance metrics}\label{Table:2}
		\centering
		\begin{tabular}{|l|l|}
			\hline
			Notation & Description\\ 
			\hline 
			$ \bP_{\Psi}^0(\SIR>\tau,\text{SL})$  & Sidelink coverage probability \\
			\hline 
			$ \bP_{\Psi}^0(\SIR>\tau,\text{DL})$ & Downlink coverage probability \\ 
			\hline 
			$ \bP_{\Psi}^0(0\in\cD(t)) $& Sidelink association probability\\ 
			\hline
			$ \bP_{\Psi}^0(0\notin\cD(t)) $& Downlink association probability\\ 
			\hline
			$ \cT $ & Downlink effective rate per user \\  
			\hline  
		\end{tabular}
	\end{table}
\end{center}

\section{Result: Coverage Probability}
This section analyzes the reliability of the sidelink and downlink communications. We derive the coverage probabilities of the typical user. Since the typical user at the origin is associated with either a vehicle or with a base station, the interference and SIR seen by the typical user varies w.r.t. the association type. For instance, when the typical user is associated with a vehicle, the origin is contained in a disk (sidelink association). In other words, there exists a vehicle within a distance $ \rho $ that covers the origin. On the other hand, when the typical user is associated with a base station, the origin is not contained in any disk.

\subsection{User Association}\label{S:3-A}
Here, we focus on the derivation of the probability that the typical user is associated with a vehicle or with a base station. The vehicle association region depends on time as shown in Eq. \eqref{setD}. However, we will show below that the association probability is not a function of time. 
\begin{theorem}\label{Theorem:1}  
	The probability that the typical user is associated with a vehicle is 
	\begin{equation}\label{eq:T1-1}
		\bP_{\Psi}^0(\text{SL}):=\bP_{\Psi}^0(0\in\cD)=1-\exp\left(-2\lambda_l\int_0^{\rho}1-\exp(-2\mu\sqrt{\rho^2-u^2})\diff u\right).
	\end{equation}
	The probability that the typical user is associated with a base station is 
	\begin{equation}\label{eq:T1-2}
				\bP_{\Psi}^0(\text{DL}):=\bP_{\Psi}^0(0\notin\cD)=\exp\left(-2\lambda_l\int_0^{\rho}1-\exp(-2\mu\sqrt{\rho^2-u^2})\diff u\right).
	\end{equation}
	Note that both Eqs are not functions of time. 
\end{theorem}

\begin{figure}
	\centering
	\includegraphics[width=0.7\linewidth]{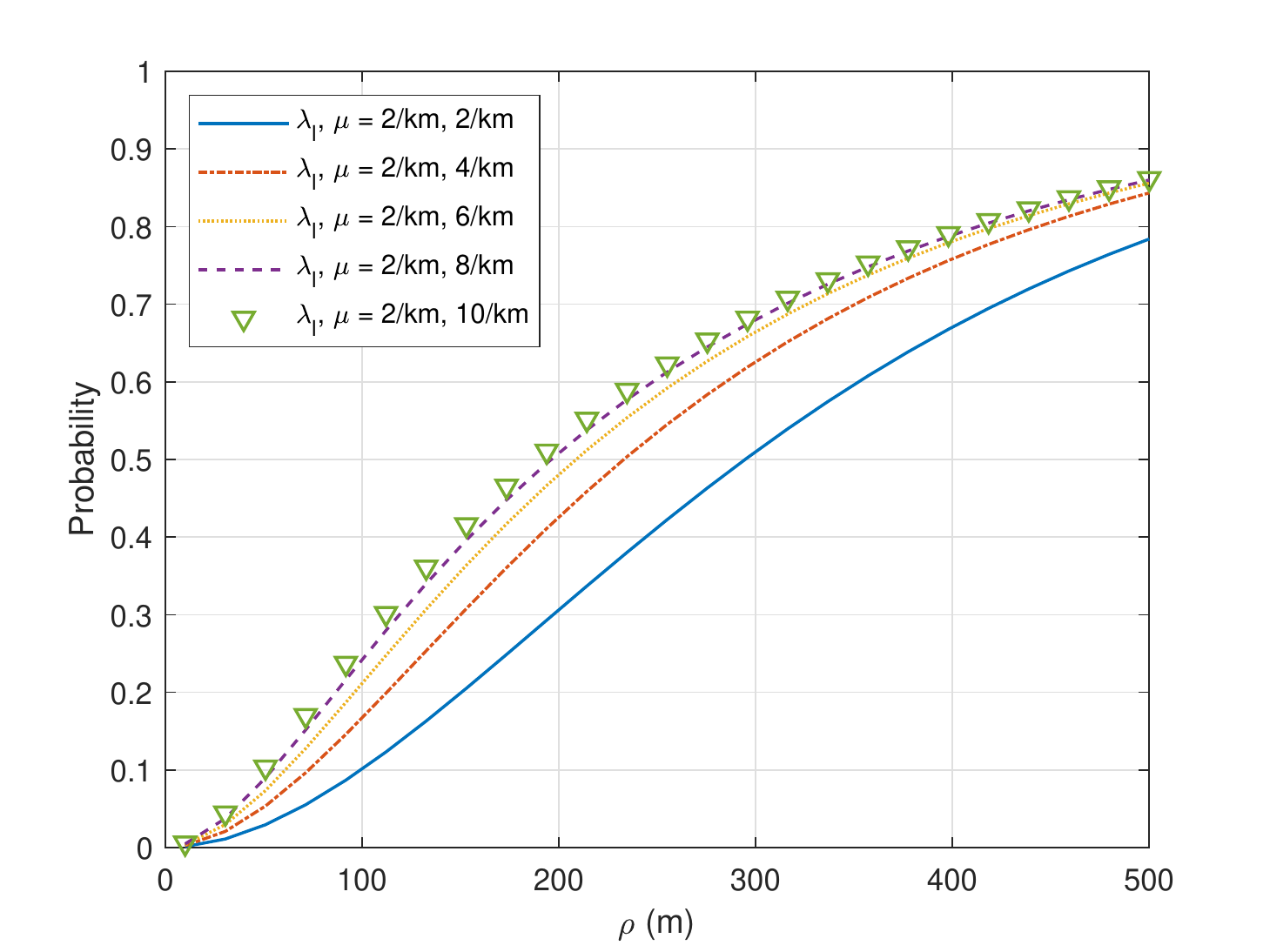}
	\caption{This illustrates the probability that the typical user is associated with sidelink broadcasts.}
	\label{fig:association2}
\end{figure}

\begin{IEEEproof}
	The probability that the typical user is associated with a vehicle is the same as the probability that there is at least one vehicle at a distance less than $ \rho $ from the origin. Hence,
	\begin{align}
		\bP_{\Psi}^0(0\in\cD(t))&=1-\bP_{\Psi}^0(\|X_i\|>\rho, \forall X_i\in\Phi_{v;t})\nnb\\
		&=1-\bP(\|X_i\|>\rho, \forall X_i\in\Phi_{v;t})\nnb\\
		&=1-\bE\left[\prod_{X_i\in\Phi_{v;t}}\ind_{\|X_i\|>\rho}\right],\label{eq:T1-step1}
	\end{align}
	where we used the fact that $ \Psi $ and $ \Phi_{v;t} $ are independent. 
	$ \Phi_{v;t} $ is the collection of point processes conditionally on $ \Phi_l. $ Thus, the vehicle point process can be written as follows:
	\begin{align}\label{eq:vehiclepp}
		\Phi_{v;t}=\sum_{(r_i,\theta_i)\in\Phi_l}\phi(r_i,\theta_i;t)
				  =\sum_{r_i\in\Phi_l}\phi(r_i;t),
	\end{align}
	where $ \phi({r_i,\theta;t}) $ denotes the Poisson point process on the line parameterized by $ (r_i,\theta_i) $ at time $ t. $ For simplicity, we also denote the Poisson point process by $ \phi(r_i;t) $. At time zero, $ \phi({r_i;0}) $ is given by a one-dimensional Poisson point process with intensity $ \mu $. Based on the considered mobility model, every vehicle randomly chooses its moving direction on its line $ (r_i,\theta_i) $ at time $ 0 $. At time $ t, $ the Poisson point process with intensity $ \mu/2 $ is translated by $ vt $ in one direction; similarly, the Poisson point process with intensity $ \mu/2 $ is translated by $ vt $ in the other direction. Based on the displacement theorem \cite{baccelli2010stochastic}, the point process $ \phi ({r_i;t}) $ at any time $ t>0 $ is a Poisson point process with intensity $ \mu $. Finally, the underlying line process $ \Phi_l $ is time-invariant and we have that the vehicle point process at time $ t $ and the vehicle point process at time $ 0 $ are equal in distribution; $\Phi_{v;t} \stackrel{d}{=}\Phi_{v;0}. $
	As a result, we write the vehicle point process as
\begin{equation}
	\Phi_v=\sum_{r_i\in\Phi_l}\phi(r_i).\label{eq:location of vehicle}
\end{equation}
	\par Due to the stationary and time-invariant properties of $ \Phi_{v;t}, $ the expectation in Eq. \eqref{eq:T1-step1} is 
	\begin{align}
		\bE_{\Phi_{v;t}}\left[\prod_{X_i\in\Phi_{v;t}}\ind_{\|X_i\|>\rho}\right]&=\bE_{\Phi_{v}}\left[\prod_{X_i\in\Phi_{v}}\ind_{\|X_i\|>\rho}\right]\nnb\\
		&=\bE_{\Phi_l}\left[\bE_{\phi(r_i)}\left[\left.\prod_{X_j\in\phi{(r_i)}}\ind_{\|X_j\|>\rho}\right|\Phi_l\right]\right]\nnb\\
		&=\bE_{\Phi_l}\left[\prod_{r_i\in\Phi_l}\bE_{\phi(r_i)}\left[\left.\prod_{X_j\in\phi(r_i)}\ind_{\|X_j\|>\rho}\right|\Phi_l\right]\right]\nnb\\
		&=\bE_{\Phi_l}\left[\prod_{r_i\in\Phi_l}\bE_{\phi(0)}\left[\left.\prod_{Z_j\in\phi(0)}\ind_{\|r\vec{a}+Z_j\vec{a}^\perp\|>\rho}\right|\Phi_l\right]\right]\nnb\\
&=\bE_{\Phi_l}\left[\prod_{r_i\in\Phi_l}\bE_{\phi(0)}\left[\left.\prod_{Z\in\phi(0)}\ind_{\|Z\|>\sqrt{\rho^2-r^2}}\right|\Phi_l\right]\right],
	\end{align}
	where we replace the Poisson point process $ \phi ({r_i}) $ with the Poisson point process $ \phi (0,\theta_i)$ using its displacement $ r_i. $  In the above, $ \vec{a} $ is the vector from the origin to the line with index $ i $. 
	Then, we have 
	\begin{equation}
		\bE_{\phi(0)}\left[\prod_{Z_j\in\phi(0)}\ind_{\|Z_j\|>\sqrt{\rho^2-r^2}}\right]=\exp(-2\mu\sqrt{\rho^2-r^2}).
	\end{equation}
	As a result, we have 
	\begin{align}
	\bP_{\Psi}^0(0\in\cD(t))&=1-\bE_{\Phi_l}\left[\prod_{X_i\in\Phi_{v}}\ind_{\{\|X_i\|>\rho\}}\right]\nnb\\
	&=1-\bE_{\Phi_l}\left[\prod_{r_i\in\Phi_l}^{|r_i|<\rho}\exp(-2\mu\sqrt{\rho^2-r^2})\right]\nnb\\
		&=1-\exp\left(-2\lambda_l\int_{0}^{\rho}1-\exp(-2\mu\sqrt{\rho^2-r^2})\diff r\right),
	\end{align}
	where we use the probability generating functional of the cylinder Poisson point process with intensity $ \lambda_l/\pi $\cite{daley2007introduction}. We use $ \bP(0\notin\cD(t))=1-\bP(0\in\cD) $ to complete the proof. 
\end{IEEEproof}
Fig. \ref{fig:association2} plots the probability that the typical user at the origin is associated with a vehicle. The figure shows the way that the probability of vehicle association increases as the density $ \mu $ increases. It also shows that the linear density $ \mu $ yields a diminishing return of the association probability. In the asymptotic regime where the linear density $ \mu $ is very high, we have Eq. \eqref{eq:T1-1} as $ \bP_{\Psi}^0(\text{SL})= 1-e^{-2\lambda_l\rho} $. This formula corresponds to the probability that the disk of radius $ \rho $ centered at the origin intersects with the Poisson line process of intensity $ \lambda_l. $ 

\begin{remark}
		We can interpret Theorem \ref{Theorem:1} in an ergodic sense \cite{baccelli2013elements}. For a given geometry of roads, base stations, and sidelink broadcast distance $ \rho, $ Eq. \eqref{eq:T1-1} gives the probability that the typical user is covered by the vehicle broadcast. In other words, it gives the instantaneous fraction of users that are receiving the sidelink broadcast. In addition, in an ergodic sense, the sidelink association probability Eq. \eqref{eq:T1-1} also corresponds to the fraction of the time that a typical user is associated with a sidelink vehicle. Similarly, the downlink association probability corresponds to the fraction of time that the user is associated with a downlink base station.
\end{remark}

\subsection{SIR Coverage of the Typical Cellular Downlink}\label{S:IV-B}
This section derives the SIR coverage probability of downlink.
\begin{theorem}\label{Theorem:3}
	The downlink coverage probability of the typical user $ \bP(\SIR>\tau, 0\notin \cD) $ is 
	\begin{equation}
	\int_{0}^\infty K_0(x)\exp(-\pi\lambda_bK_1(x)-2\lambda_lK_2(x)-2\lambda_lK_3(x))\diff x,\label{eq:Theorem2}
	\end{equation} 
	where 
	\begin{align*}
	&K_0(x)=2\pi\lambda_b x,\\
	&K_1(x)= x^2 + 2\int_{x}^\infty\frac{\tau x^\alpha  r^{1-\alpha}}{1+\tau x^\alpha r^{-\alpha}}\diff r,\\
	&K_2(x)= \int_{0}^\rho 1-\exp\left(-2\mu\sqrt{\rho^2-r^2}-2\mu\int_{\sqrt{\rho^2-r^2}}^{\infty}\frac{\tau x^\alpha \eta{(r^2+t^2)}^{-\frac{\alpha}{2}}}{1+\tau x^\alpha \eta{(r^2+u^2)}^{-\frac{\alpha}{2}}}\diff u\right)\diff r,\\
	&K_3(x)= \int_{\rho}^\infty 1-\exp\left(-2\mu\int_{0}^{\infty}\frac{\tau x^\alpha \eta{(r^2+u^2)}^{-\frac{\alpha}{2}}}{1+\tau x^\alpha \eta{(r^2+u^2)}^{-\frac{\alpha}{2}}}\diff u\right)\diff r.
	\end{align*}
	We define the transmit power ratio $ \eta \stackrel{d}{=}p_v/p_b $
\end{theorem}
\begin{IEEEproof}
	The probability that the typical user is not in $ \cD $ and its SIR is greater than $ \tau $ is 
	\begin{align}
	&\bP_{\Psi}^0(\SIR>\tau,0\notin\cD)\nnb\\
	&=\bE_{\Phi_l}\left[\bP\left(\left.H>\tau {\|X_0\|}^{\alpha}p_v^{-1}I 
	,0\notin\cD\right|\Phi_l\right)\right]\nnb\\
	&=\bE_{\Phi_l}\int_0^{\infty}\bP\left(\left.H>\tau x^{\alpha} p_b^{-1}I 
	,\right|\|X_0\|=x,0\notin\cD,\Phi_l\right)f(x)\diff x\nnb\\
	&=\bE_{\Phi_l}\int_0^{\infty}\bE\left[\exp(-\tau x^{\alpha} p_b^{-1}I 
)|\|X_0\|=x,0\notin\cD,\Phi_l\right]f(x)\diff x\label{23},
	\end{align}
	where $ I= \sum_{X_i\in\Phi_b\setminus B_0(x)+\Phi_v\setminus B_0(\rho)}p_iH_i\|X_i\|^{-\alpha} $ is the shot-noise process seen by the typical user and $ \|X_0\| $ is the distance from the typical user to its nearest base station. In Eq. \eqref{23}, $ f(x) $ is the density function of $ \|X_0\| $ on the event $ 0\notin\cD, $ conditional on $ \Phi_l $, namely, 
	\begin{align}
	f(x)\diff x&=\bP(x \leq \|X_0\|\leq x+\diff x, 0\notin \cD|\Phi_l)\nnb\\
	&= \bE\left[\left.\sum_{X_i\in\Phi_b}\ind_{x\leq \|X_i\|\leq x+\diff x}\ind_{\Phi_b(B_0(\|X_i\|))=0}\ind_{\Phi_v(B_0(\rho))=0}\right|\Phi_l\right]\nnb\\
	&=\int_{0}^{2\pi} \ind_{x\leq z \leq x+\diff x}\bP_z^0(\Phi_b(B_0(z))=0)\bP_z^0(\Phi_v(B_0(\rho))=0|\Phi_l)\lambda_bx\diff x\diff \theta\nnb\\
	&=2\pi\lambda_b x \diff xe^{-\pi\lambda_b x^2 }\prod_{r_i\in\Phi_l}^{|r_i|<\rho}\bE_{\phi(r_i)}\left[\prod_{X_j\in\phi(r_i)}\ind_{\|X_j\|>\rho}\right]\nnb\\
	&=2\pi\lambda_bx\diff xe^{-\pi\lambda_b x^2 }\prod_{r_i\in\Phi_l}^{|r_i|<\rho}\bE_{\phi(r_i)}\left[\prod_{Y_j\in\phi(0)}\ind_{\|Y_j\|>\sqrt{\rho^2-r_i^2}}\right]\nnb\\
	&=2\pi\lambda_bx\diff xe^{-\pi\lambda_b x^2 }\prod_{r_i\in\Phi_l}^{|r_i|<\rho}e^{-2\mu\sqrt{\rho^2-r_i^2}}\label{23-1}.
	\end{align}
	The integrand of Eq. \eqref{23} is the Laplace transform of the interference from vehicles outside of the disk $ B_0(\rho) $ and from base station outside of the disk $ B_0(x) $ conditional on $ \|X_0\|=x $ and $ \Phi_l $. Specifically, the integrand is 
	\begin{align}
		\bE\left[\left.\exp\left(-\tau x^\alpha p_b^{-1}\hspace{-.5cm}\sum_{X_k\in\Phi_b\setminus B_0(x)}\hspace{-.5cm}p_bH_k\|X_k\|^{-\alpha} - \tau x^\alpha p_b^{-1}\hspace{-.5cm}\sum_{X_l\in\Phi_v\setminus B_0(\rho)}\hspace{-.5cm} p_vH_l\|X_l\|^{-\alpha}\right)\right|\|X_0\|=x ,0\notin\cD,\Phi_l\right].\nnb
	\end{align}
	\par By using the independence of $ \Phi_l $ and $ \Phi_b $ and by applying the Laplace transform of the interference from cellular base stations outside of the disk of radius $ x $ \cite{baccelli2010stochastic}, we obtain 
	\begin{align}
				\bE\left[\left.\exp\left(-\tau x^\alpha \hspace{-.5cm}\sum_{X_k\in\Phi_b\setminus B_0(x)}\hspace{-.5cm}H_k\|X_k\|^{-\alpha} \right)\right|\|X_0\|=x,0\notin\cD,\Phi_l\right] = e^{-2\pi\lambda_b\int_{x}^\infty\frac{\tau x^\alpha  r^{1-\alpha}}{1+\tau x^\alpha r^{-\alpha}}\diff r}.\label{eq:T2-laplace-3}
	\end{align}
	Using the independence and the power ratio $ \eta \stackrel{\triangle}{=} p_v/p_b $, we have 
	\begin{align}
		\bE\left[\left.\exp\left(-\tau x^\alpha \eta \hspace{-.5cm}\sum_{X_l\in\Phi_v\setminus B_0(\rho)}\hspace{-.5cm}H_l\|X_l\|^{-\alpha} \right)\right|0\notin\cD,\Phi_l\right]=&\prod_{r_i\in\Phi_l}^{|r_i|<\rho}\bE_{\phi(r_i)}\left[\prod_{Y_m\in\phi(r_i)}^{\|Y_m\|>\rho}e^{-\tau x^\alpha\eta \|Y_m\|^{-\alpha}}\right]\nnb\\
		&\times\prod_{r_i\in\Phi_l}^{|r_i|>\rho}\bE_{\phi(r_i)}\left[\prod_{Y_n\in\phi(r_i)}e^{-\tau x^\alpha\eta \|Y_n\|^{-\alpha}}\right].
	\end{align}
	In distribution, the points $ \{Y_m\}_m $ of the Poisson point process $ \phi(r_i) $ can be represented in terms of the points $ \{Z_m\}_m $ of the Poisson point process $ \phi(0) $ through $ \{\|Y_m\|= (Z_m^2+r_i^2)^{\frac{1}{2}}\}_{m}. $ Hence, we have 
	\begin{align}
		\prod_{r_i\in\Phi_l}^{|r_i|<\rho}\bE_{\phi(r_i)}\left[\left.\prod_{Y_m\in\phi(r_i)}^{\|Y_m\|>\rho}e^{-\tau x^\alpha\eta \|Y_m\|^{-\alpha}}\right|\Phi_l\right]
		&= \prod_{r_i\in\Phi_l}^{|r_i|<\rho|}\bE_{\phi(0)}\left[\prod_{Z_m\in\phi(0)}^{|Z_m|>\sqrt{\rho^2-r_i^2}}\frac{1}{1+\tau x^{\alpha}\eta {(r_i^2+Z_m^2)}^{-\frac{\alpha}{2}}}\right]\nnb\\
		&=\prod_{r_i\in\Phi_l}^{|r_i|<\rho|}\exp\left(-2\mu\int_{\sqrt{\rho^2-r_i^2}}^{\infty}\frac{\tau x^\alpha \eta{(r_i^2+u^2)}^{-\frac{\alpha}{2}}}{1+\tau x^\alpha \eta{(r_i^2+u^2)}^{-\frac{\alpha}{2}}}\diff u\right).\label{25}
	\end{align}
	Similarly, we have
	\begin{align}
		\prod_{r_i\in\Phi_l}^{|r_i|>\rho}\bE_{\phi(r_i)}\left[\left.\prod_{Y_n\in\phi(r_i)}^{\|Y_n\|>\rho}e^{-\tau x^\alpha\eta \|Y_n\|^{-\alpha}}\right|\Phi_l\right]
		&=\prod_{r_i\in\Phi_l}^{|r_i|>\rho}\exp\left(-2\mu\int_{0}^{\infty}\frac{\tau x^\alpha \eta{(r_i^2+u^2)}^{-\frac{\alpha}{2}}}{1+\tau x^\alpha \eta{(r_i^2+u^2)}^{-\frac{\alpha}{2}}}\diff u\right)\label{26}.
	\end{align}
	Then, from Eqs. \eqref{23-1}--\eqref{26}, the coverage probability is given by 
	\begin{align}
		2\pi\lambda_b\int_{0}^{\infty}xe^{-\pi\lambda_bx^2-2\pi\lambda_b\int_x^{\infty}\frac{\tau x^\alpha u^{1-\alpha}}{1+\tau x^\alpha u^{-\alpha}\diff u} x^2}\bE\left[\prod_{r_i\in\Phi_l}^{|r_i|<\rho}f(r_i,x)\prod_{r_i\in\Phi_l}^{|r_i|>\rho}g(r_i,x)\right]\diff x \nnb,
	\end{align}
	where the functions $ f(r_i,x) $ and $ g(r_i,x) $ are given in Eq. \eqref{25} and \eqref{26}, respectively. Finally, we obtain the final result by using the Laplace transform of the Poisson line process. 
\end{IEEEproof}

\begin{figure}
	\centering
	\includegraphics[width=0.8\linewidth]{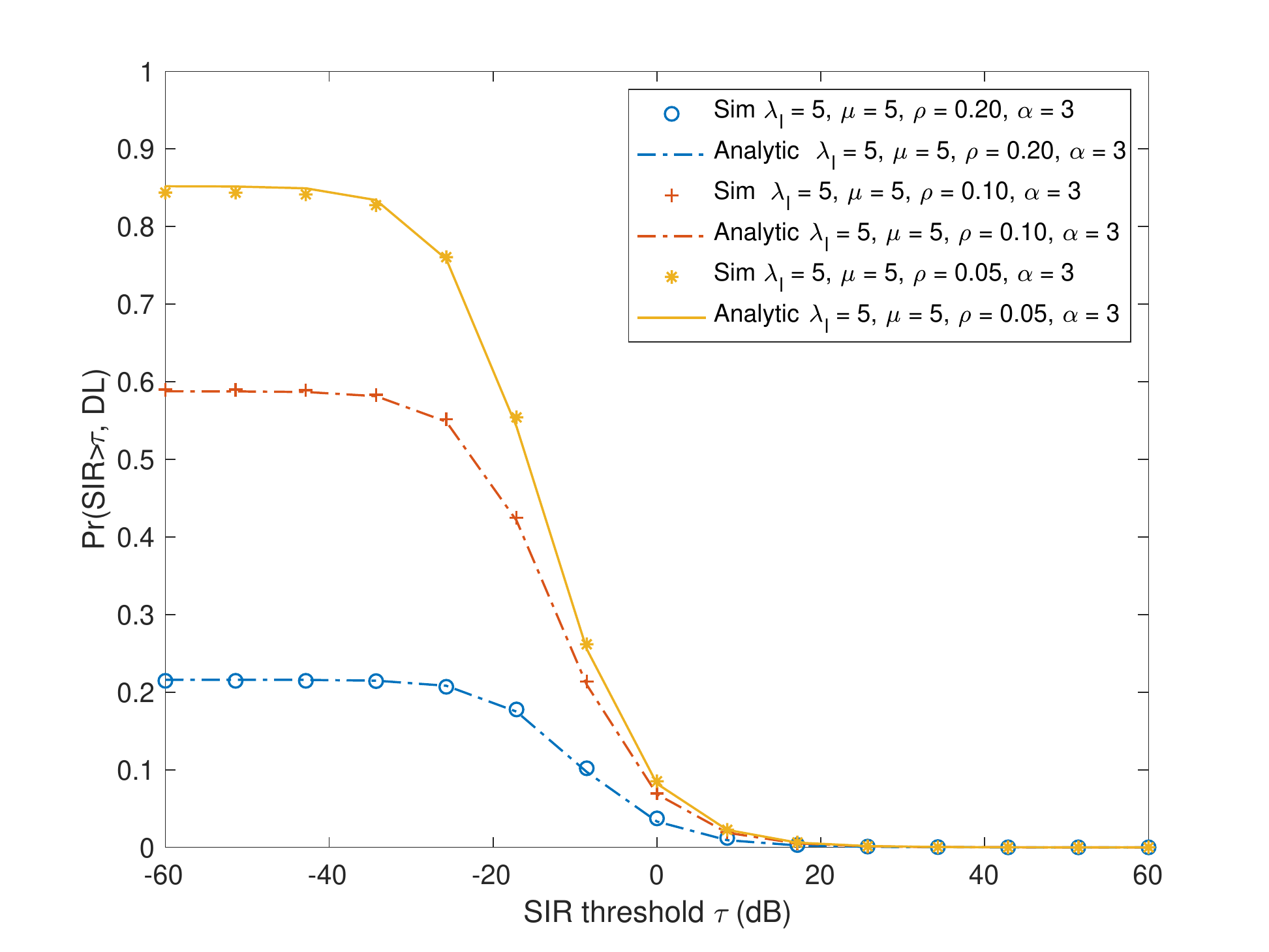}
	\caption{Comparison of the simulated joint downlink coverage probability and the analytic formula derived in Theorem \ref{Theorem:3}. This figure confirms that the analytic formula matches the simulation results. We take $ \lambda_b =5 /\text{km}^2$. The densities $ \lambda_l,\mu $ are per kilometer and $ \rho $ is in kilometer.}
	\label{fig:DL_coverage}
\end{figure}

\begin{figure}
	\centering
	\includegraphics[width=0.8\linewidth]{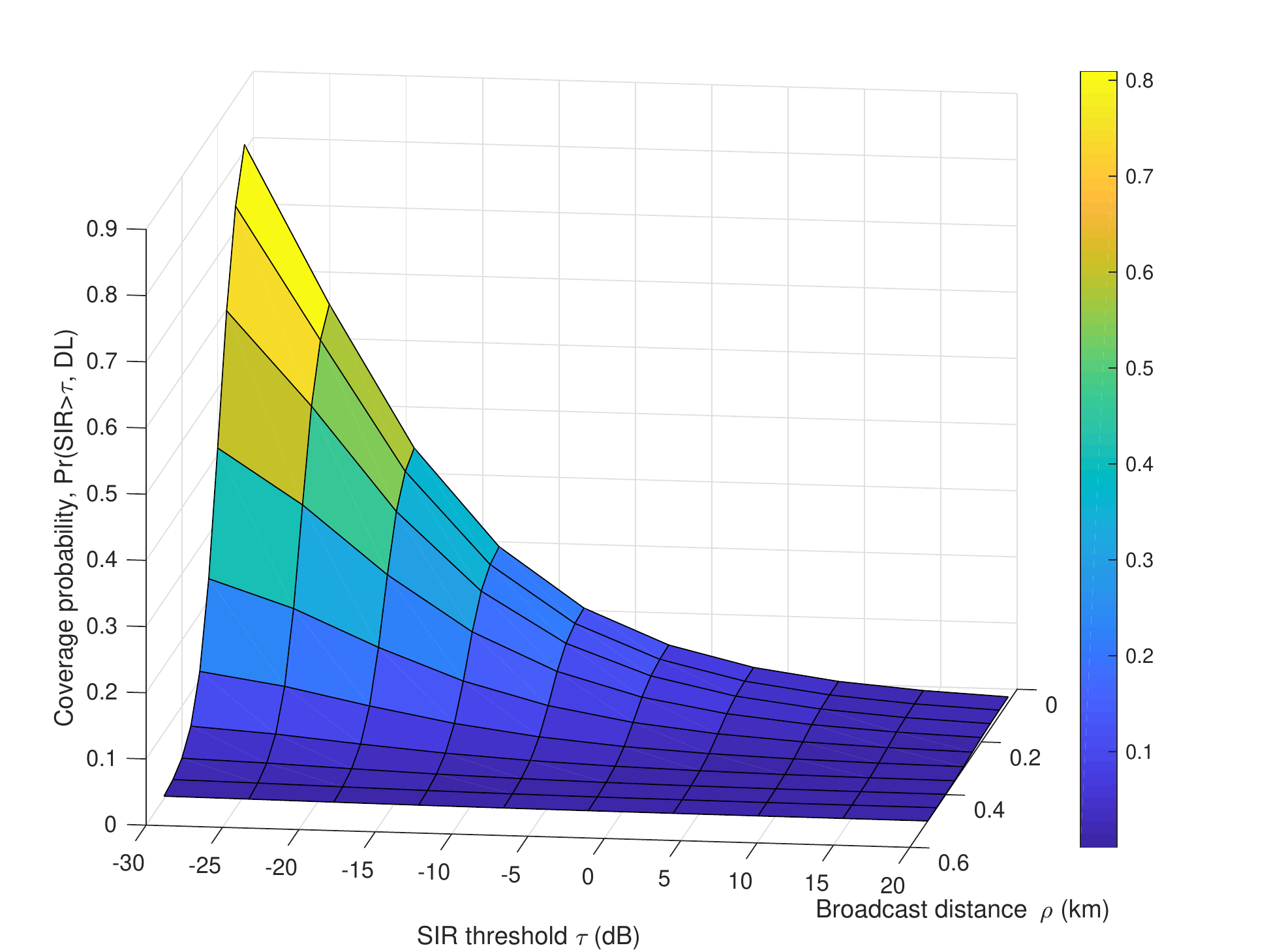}
	\caption{We consider $ \lambda_b =5 /\text{km}^2, \lambda_l=5/\text{km}, \mu = 5/\text{km}$.}
	\label{fig:DLmesh}
\end{figure}

Fig. \ref{fig:DL_coverage} illustrates the coverage probability of the typical user, obtained by Monte Carlo simulations and by Theorem \ref{Theorem:3}.  Fig. \ref{fig:DLmesh} illustrates the coverage probability of the typical user w.r.t. the safety message range $ \rho $ and SIR coverage threshold $ \tau $. It shows how the SIR coverage decreases when $  \tau $ or $ \rho $ increase.  

\subsection{SIR Coverage of the Typical Sidelink}\label{S:IV-A}
This section derives the SIR coverage probability of sidelink.
\begin{theorem}\label{Theorem:2}
	The sidelink coverage probability of the typical user $ \bP_{\Psi}^0(\SIR>\tau,0\in\cD) $ is 
	\begin{align}\label{eq:T3}
\int_0^{\rho}L_0(x)\exp(-L_1(x)-L_2(x)-L_3(x))\diff x,
	\end{align}
	where 
	\begin{align*}
		L_0(x) &= \int_0^{x}\frac{4\lambda_l\mu x\exp\left({-2\mu\sqrt{x^2-r^2}-2\mu\int_{\sqrt{x^2-r^2}}^{\infty}\frac{\tau x^\alpha (u^2+r^2)^{-\frac{\alpha}{2}}}{1+\tau x^\alpha (u^2+r^2)^{-\frac{\alpha}{2}}}}\diff u\right)}{\sqrt{x^2-r^2}}\diff r\\
		L_1(x)& = 2\pi\lambda_b\int_0^{\infty}\frac{\tau \eta^{-1} x^\alpha u^{1-\alpha}}{1+\tau^{-1}\eta x^\alpha u^{-\alpha}}\diff u\\
		L_2(x)&= 2\lambda_l\int_{0}^{x}1-\exp\left(-2\mu\sqrt{x^2-u^2}-2\mu\int_{\sqrt{x^2-u^2}}^{\infty}\frac{\tau x^\alpha(u^2+v^2)^{-\frac{\alpha}{2}}}{1+\tau x^\alpha(u^2+v^2)^{-\frac{\alpha}{2}}}\diff v\right)\diff u\\
		L_3(x)&=2\lambda_l\int_{x}^{\infty}1-\exp\left(-2\mu\int_{0}^{\infty}\frac{\tau x^\alpha(u^2+v^2)^{-\frac{\alpha}{2}}}{1+\tau x^\alpha(u^2+v^2)^{-\frac{\alpha}{2}}}\diff v\right)\diff u.
	\end{align*}
\end{theorem}
\begin{IEEEproof}
	The coverage probability is given by 
	\begin{align}
		\bP(\SIR>\tau,\text{SL})&=\bP(\SIR>\tau,0\in\cD)\nnb\\
		&=\bE\left[\bE\left[\ind_{\SIR>\tau,0\in\cD}|\Phi_l\right]\right]\nnb\\
		&=\bE\left[\left.\bE\left[\sum_{r_i\in\Phi_l}\ind_{X_\star \in r_i}\ind_{\|X_\star\| =x} \ind_{H>p_v^{-1}\tau I x^\alpha}\right|\Phi_l\right]\right]\nnb\\
		&=\bE\left[\int_{0}^\rho\sum_{r_i\in\Phi_l}^{r_i<x}f(r_i,x)\bE\left[\left.e^{-p_v^{-1}\tau I x^\alpha}\right|X_i\in r_i, \|X_\star\|=x , \Phi_l\right] \diff x \right]\label{24},
	\end{align}
	where the conditional density function $ f(r_i,x) $ is given by 
		\begin{align}
	f(r_i,x) &= \partial_x\bP(x \leq \|X_\star\| ,X_\star \in r_i|\Phi_l)\nnb\\
	&=\partial_x(1-\bP(\phi(r_i)(B_0(x))=0|\Phi_l))\prod_{r_j\in\Phi_l, r_j\neq r_i}^{|r_j|<x}\bP(\phi(r_j)  B_0(x)=0|\Phi_l)\nnb\\
	&=\partial_x(1-e^{-2\mu\sqrt{x^2-r_i^2}})\prod_{r_j\in\Phi_l, r_j\neq r_i}^{|r_j|<x}e^{-2\mu\sqrt{x^2-r_j^2}}\nnb\\
	&=\frac{2\mu x e^{-2\mu \sqrt{x^2-r_i^2}}}{\sqrt{x^2-r_i^2}}\prod_{r_j\in\Phi_l, r_j\neq r_i}^{|r_j|\leq x}e^{-2\mu\sqrt{x^2-r_j^2}}\label{eq25}.
	\end{align}
		The first term is the probability that there is a point of $ \phi(r_i) $ at distance $ x $ from the origin. The second term is the probability that there exists no point of $ \phi(r_j) $, for any $ r_j\in\Phi_l $ such that $ |r_j|<x $.
		\par 
	
	Conditional on $ X_\star= x, X_\star\in r_i,$ and $\Phi_l,  $ the Laplace transform of the interference is 
	\begin{align}
	\bE[e^{-p_v^{-1}\tau I x^\alpha}|\|X_\star\|=x, X_\star \in r_i, \Phi_l] =&\bE_{}\left[\prod_{Y_i\in\Phi_b}e^{-\tau \eta^{-1} x^\alpha H \|Y_i\|^{-\alpha}}\right]\nnb\\ &\times\bE_{\phi(r_i)}\left[\prod_{X_j\in\phi(r_i)}^{\|X_j\|>x}\bE_H\left[e^{-\tau x^\alpha H \|X_j\|^{-\alpha}}\right]\right]\nnb\\
	&\times\prod_{r_j\in\Phi_l,r_j\neq r_i}\bE_{\phi(r_j)}\left[\prod_{X_j\in\phi(r_j)}^{\|X_j\|>x}\bE_H\left[e^{-\tau x^{\alpha}H \|X_j\|^{-\alpha}}\right]\right]\nnb\\
	=&\exp\left(-2\lambda_b\pi\int_0^{\infty}\frac{\tau\eta^{-1} x^{\alpha}u^{1-\alpha}}{1+\tau \eta^{-1} x^{\alpha}u^{-\alpha}}\diff u\right)\nnb\\
	&\times \exp\left(-2\mu\int_{\sqrt{x^2-r_i^2}}^{\infty}\frac{\tau x^\alpha \left(r_i^2+u^2\right)^{-\frac{\alpha}{2}}}{1+\tau x^\alpha (r_i^2+u^2)^{-\frac{\alpha}{2}}}\diff u\right)\nnb\\
	&\times \prod_{r_j\in\Phi_l,r_j\neq r_i}^{|r_j|<x}\exp\left(-2\mu\int_{\sqrt{x^2-r_j^2}}^{\infty}\frac{\tau x^\alpha \left(r_j^2+u^2\right)^{-\frac{\alpha}{2}}}{1+\tau x^\alpha (r_j^2+u^2)^{-\frac{\alpha}{2}}}\diff u\right)\nnb\\
	&\times \prod_{r_j\in\Phi_l, r_j\neq r_i}^{|r_j|>x} \exp\left(-2\mu\int_{0}^{\infty}\frac{\tau x^\alpha \left(r_j^2+u^2\right)^{-\frac{\alpha}{2}}}{1+\tau x^\alpha (r_j^2+u^2)^{-\frac{\alpha}{2}}}\diff u\right)\label{e24}.
	\end{align}
On the right hand side of Eq. \eqref{e24}, the conditional expectation w.r.t. $ \Phi_l $ are ommitted. 
%

	
	\par By incorporating Eqs. \eqref{eq25} and \eqref{e24} to Eq. \eqref{24}, we can write the coverage probability as 
	\begin{align}
		&\int_{0}^\rho \bE_{\Phi_l}\left[\sum_{r_i\in\Phi_l}^{|r_i|<x}g_1(r_i,x)\prod_{r_j\in\Phi_l,r_j\neq r_i}^{|r_j|<x}g_2(r_j,x)\prod_{r_j\in\Phi_l, r_j\neq r_i}^{|r_j|>x}g_3(r_j,x)\right] \xi(x)\diff x\label{eq:27},
	\end{align}
	where the integrand functions are given by 
	\begin{align*}
		g_1(r_i,x)&=\frac{2\mu x e^{-2\mu \sqrt{x^2-r_i^2}}}{\sqrt{x^2-r_i^2}}\exp\left(-2\mu\int_{\sqrt{x^2-r_i^2}}^{\infty}\frac{\tau x^\alpha \left(r_i^2+v^2\right)^{-\frac{\alpha}{2}}}{1+\tau x^\alpha (r_i^2+v^2)^{-\frac{\alpha}{2}}}\diff v\right),\\
		g_2(r_j,x)&=e^{-2\mu\sqrt{x^2-r_j^2}}\exp\left(-2\mu\int_{\sqrt{x^2-r_j^2}}^{\infty}\frac{\tau x^\alpha \left(r_j^2+v^2\right)^{-\frac{\alpha}{2}}}{1+\tau x^\alpha (r_j^2+v^2)^{-\frac{\alpha}{2}}}\diff v\right),\\
		g_3(r_j,x)&=\exp\left(-2\mu\int_{0}^{\infty}\frac{\tau x^\alpha \left(r_j^2+v^2\right)^{-\frac{\alpha}{2}}}{1+\tau x^\alpha (r_j^2+v^2)^{-\frac{\alpha}{2}}}\diff v\right),\\
		\xi(x) & = \exp\left(-2\pi\lambda_b\int_0^{\infty}\frac{\tau\eta^{-1} x^{\alpha}u^{1-\alpha}}{1+\tau \eta^{-1} x^{\alpha}u^{-\alpha}}\diff u\right)\label{311-}.
	\end{align*}
	Applying Campbell's formula in \cite{baccelli2010stochastic} on Eq. \eqref{eq:27} w.r.t. the cylinder Poisson point process gives
	\begin{align}
		\int_0^\rho \left(\int_{0}^{x}2\pi\lambda_l  g_1(r,x)\diff r \diff \theta\right)\bE_{\Phi_l}^{r}\left[\prod_{r_j\in\Phi_l, r_j\neq  r}^{|r_j|<x}g_2(r_j,x)\prod_{r_j\in\Phi_l,r_j\neq r}^{|r_j|>x}g_3(r_j,x)\right] \xi(x)\diff x,
	\end{align} 
	where $ \bE_{\Phi_l}^{r} $ is the Palm expectation of the cylinder Poisson line process. We have 
	\begin{align}
		&\bE_{\Phi_l}^{r}\left[\prod_{r_j\in\Phi_l, r_j\neq  r}^{|r_j|<x}g_2(r_j,x)\prod_{r_j\in\Phi_l,r_j\neq r}^{|r_j|>x}g_3(r_j,x)\right]\nnb\\
		&=\bE_{\Phi_l}^{!r}\left[\prod_{r_j\in\Phi_l}^{|r_j|<x}g_2(r_j,x)\prod_{r_j\in\Phi_l}^{|r_j|>x}g_3(r_j,x)\right]\nnb\\ &=\exp\left(-2\lambda_l\int_{0}^{x}1-\exp\left(-2\mu\sqrt{x^2-u^2}-2\mu\int_{\sqrt{x^2-u^2}}^{\infty}\frac{\tau x^\alpha(u^2+v^2)^{-\frac{\alpha}{2}}}{1+\tau x^\alpha(u^2+v^2)^{-\frac{\alpha}{2}}}\diff v\right)\diff u\right)\nnb\\
		&\hspace{5mm}\times \exp\left(-2\lambda_l\int_{x}^{\infty}1-\exp\left(-2\mu\int_{0}^{\infty}\frac{\tau x^\alpha(u^2+v^2)^{-\frac{\alpha}{2}}}{1+\tau x^\alpha(u^2+v^2)^{-\frac{\alpha}{2}}}\diff v\right)\diff u\right),
	\end{align} We obtain the first equation by using Slivnyak's theorem and the second equation by using the probability generating functional of the cylinder Poisson point process.
\end{IEEEproof}

\begin{figure}
	\centering
	\includegraphics[width=0.8\linewidth]{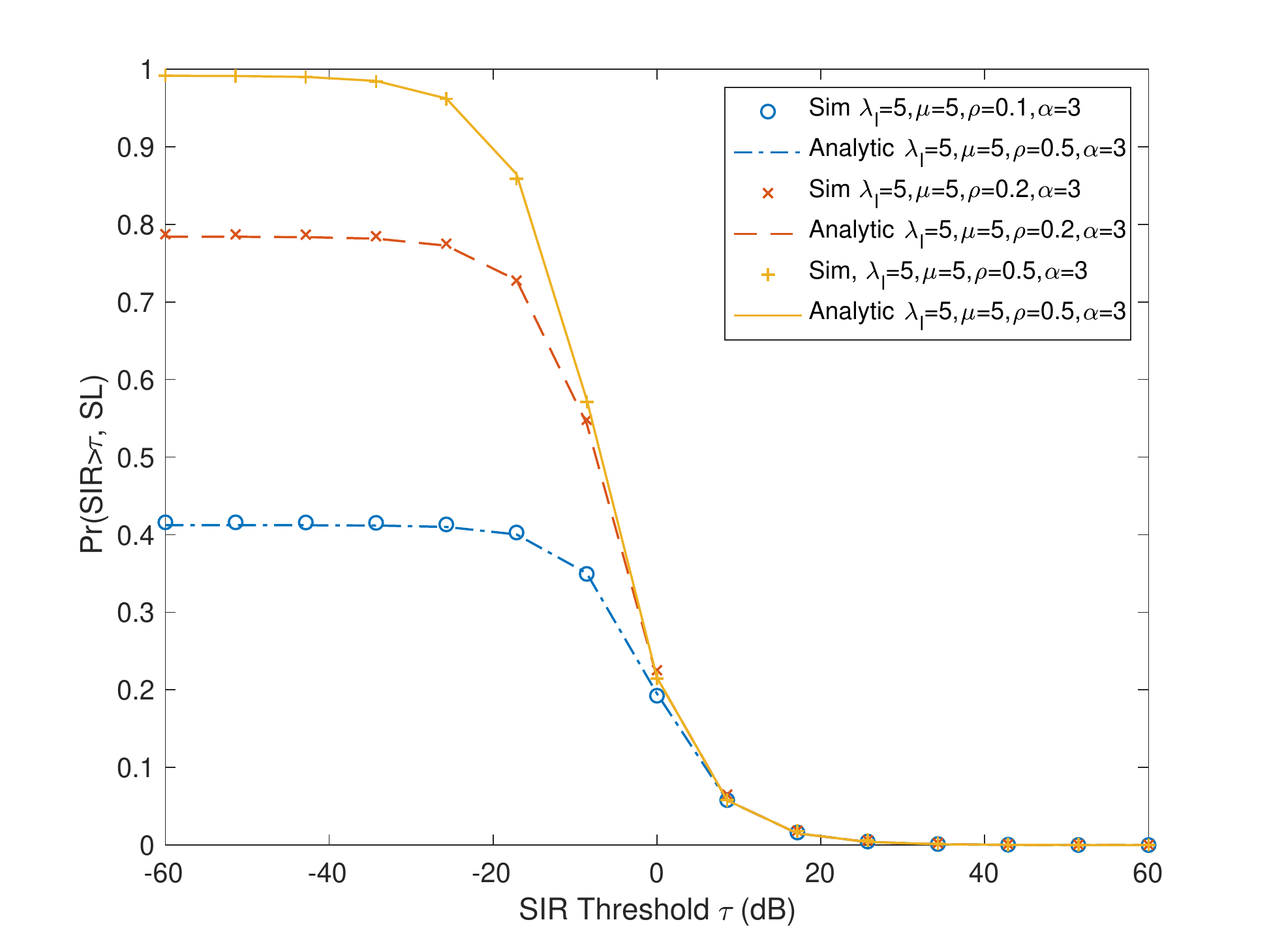}
	\caption{Comparison between the simulated sidelink coverage probability and the analytic one presented by Theorem \ref{Theorem:2}. We consider $ \lambda_b =5 /\text{km}^2$ $ \alpha=3, $ and $ \eta=1 $. The densities $ \lambda_l,\mu $ are per kilometer and $ \rho $ is in kilometer.}
	\label{fig:SL_coverage}
\end{figure}

\begin{figure}
	\centering
	\includegraphics[width=0.8\linewidth]{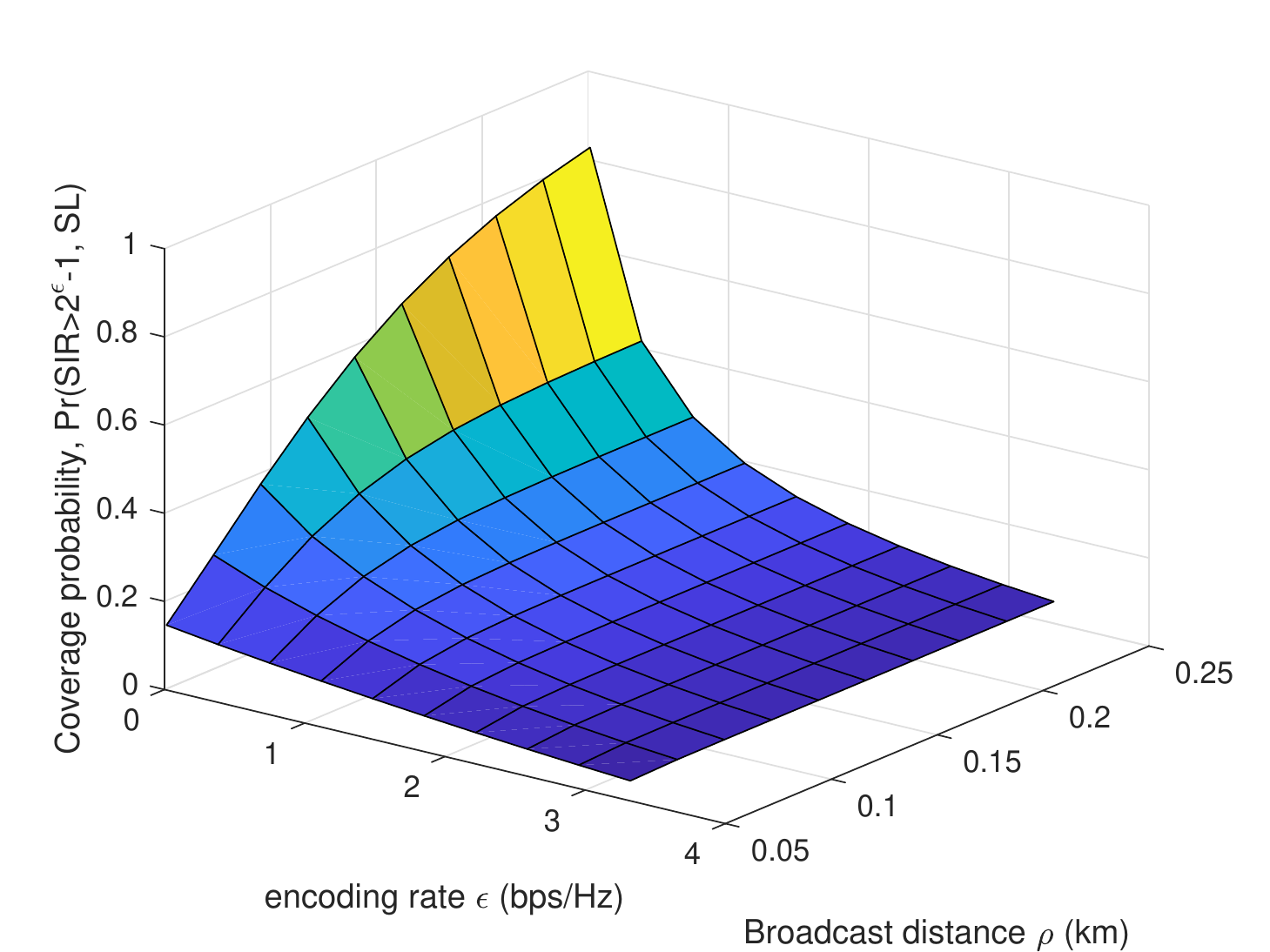}
	\caption{Sidelink coverage probability for $ \rho $ and $ \epsilon $. We consider $ \lambda_b =5 /\text{km}^2, \lambda_l=5/\text{km}, \mu = 5/\text{km}$, $ \eta=1 $, and $ \alpha=3 $. }
	\label{fig:SLmesh}
\end{figure}

Fig. \ref{fig:SL_coverage} illustrates the derived probability of the typical sidelink user, obtained by Monte Carlo simulations and by Theorem \ref{Theorem:2}, respectively. 
In the proposed network model, vehicles broadcast their safety messages at a fixed encoding rate such as $ \epsilon $ bps/Hz. The value of $ \epsilon $ determines the corresponding SIR threshold $ \tau $ at which safety messages are successfully decoded. In Fig. \ref{fig:SLmesh}, we illustrate the sidelink coverage probability as a function of the encoding threshold $ \epsilon $ (bps/Hz) and the broadcast distance $ \rho$ (km). The $ z $-value is the probability that the message with the corresponding encoding rate is successfully decoded by the typical user: $ \bP_{\Psi}^0(\SIR>2^{\epsilon}-1,\text{SL}) $.
\begin{remark}
	The sidelink coverage probability conditional on the sidelink association is the ratio of the number of vehicle-associated users whose SIRs are greater than $ \epsilon $ to the number vehicle-associated users, in a very large ball centered at the origin: 
	\begin{align}
\bP_{\Psi}^0(\SIR>2^{\epsilon}-1|\text{SL})
&=	\frac{\bP_{\Psi}^0(\SIR>2^{\epsilon}-1,\text{SL})}{\bP_{\Psi}^0(\text{SL})}\nnb\\
	&= \lim_{r\to\infty}\frac{\sum_{X_i\in\Psi\cap B_0(r)} \ind_{X_i \in \cD\SIR_{X_i}> 2^{\epsilon}-1}}{\sum_{X_i\in\Psi\cap B_{0}(r)} \ind_{X_i\in\cD}}\nnb,
	\end{align}
	where $ \SIR_X $ denotes the SIR of user $ X$ and $ \cD $ denotes the vehicle association region. The sidelink coverage probability characterizes the reliability of the vehicular broadcast safety messaging. 
\end{remark}

\section{Effective Data Rate}\label{S:4}

This section studies the effective data rate of the typical user through downlink communications. It captures the average amount of data that users will effectively receive through messages from base stations in the presense of sidelink interruptions.

\par Recall that the zero cell, denoted by $ \cV_z, $ is the Voronoi cell that contains the origin (or equivalently the typical user) under the Palm distribution of the user point process $ \Psi $. The nucleus of the zero cell corresponds to the base station that serves the typical user.  

\subsection{Fraction of Vehicle Association Region in the Zero Cell}

To derive the effective data rate, we first evaluate the expected number of downlink association users of the serving base station of the typical user: $ \bE_{\Psi}^0[\|\Psi(\cV_Z\setminus \cD(t))\|] $. We first derive the mean area of the set $ \cV_Z\setminus \cD(t) $ under the Palm distribution of $ \Psi. $ Then, the mean number of downlink users is simply given by the product of the user density $ \lambda_u $ and the mean area of the zero cell, $ \bE[\|\cV_Z\setminus \cD(t)\|]$.

\begin{lemma}[Inversion lemma]
	Suppose $ \Upsilon $ a stationary point process of intensity $ \lambda. $ For all non-negative functions $ f, $ the Palm inversion formula \cite{baccelli2010stochastic} is 
	\begin{equation}
	\bE[f(\Upsilon)]=\lambda\bE_{\Upsilon }^0\left[\int_{\bR^d}f(\Upsilon-x)\ind_{x\in\cV_0}\diff x\right].
	\end{equation}
	In addition, for any non-negative function $ g(\cdot) $ of the zero cell, we have 
	\begin{equation}
	\bE[g(\cV_Z)]=\lambda\bE_{\Upsilon}^{0}[g(\cV_0)\ell_2(\cV_0)],\label{eq:inversion}
	\end{equation}
	where $ \ell_2(A) $ is the area of the set $ A $ and $ \cV_0 $ is the typical Voronoi cell (the typical cell under the Palm distribution $ \bP_{\Upsilon} $).
\end{lemma}
\begin{proposition}
	The mean areas of the sets $ \cV_Z\cap\cD(t) $ and $ \cV_Z\setminus \cD(t) $ are given by 
	\begin{align*}
\bE[\ell_2(\cV_Z \cap \cD(t))]&=\lambda_b\bE_{\Phi_b}^0\left[(\ell_2(\cV_0))^2\right]\bP(0\in\cD)\approxeq\frac{\nu}{\lambda_b}\bP(0\in\cD),\\
\bE[\ell_2(\cV_Z \setminus \cD(t))]&=\lambda_b\bE_{\Phi_b}^0\left[(\ell_2(\cV_0))^2\right]\bP(0\notin\cD)\approxeq\frac{\nu}{\lambda_b}\bP(0\notin\cD),
\end{align*}
respectively. We use $ \nu\approxeq 1.28. $
\end{proposition}
\begin{IEEEproof}
	Let $ g(\cV_Z) $ denote the area of the set $ \cD(t) $ {in} the zero cell $ \cV_Z. $ Then, we have 
		\begin{equation*}
	g(\cV_Z)=\|\cV_Z\cap \cD(t)\|=\int_{\bR^2}\ind_{x\in\cV_Z\cap\cD(t)}\diff x.
	\end{equation*}
	Using the inversion lemma, the expected value  $ \bE[\ell_2(\cV_Z \cap \cD)] $ is given by
	\begin{align}
		\bE\left[\int \ind_{x\in\cV_Z\cap D(t)} \diff x\right]&=\lambda_b\bE_{\Phi_b}^0 \left[\ell_2(\cV_0)\int \ind_{x\in\cV_0\cap D(t)} \diff x\right]\nnb\\
		&=\lambda_b\bE_{\Phi_b}^0 \left[\int \ind_{x\in\cV_0}\diff x \int \ind_{x\in\cV_0\cap \cD(t)} \diff x\right]\nnb\\
		&\ea\lambda_b\bE_{\Phi_b}^0\left[\bE_{\Phi_{v;t}}\left[\left.\int \ind_{x\in\cV_0}\diff x\int \ind_{x\in\cV_0\cap \cD(t)} \diff x\right|\Phi_b\right]\right]\nnb\\
		&=\lambda_b\bE_{\Phi_b}^0\left[\bE_{\Phi_{v;t}}\left[\left.\int \ind_{x\in\cV_0}\diff x\int \ind_{x\in\cV_0}\ind_{x\in \cD(t)} \diff x\right|\Phi_b\right]\right]\nnb\\
		&\eb\lambda_b\bE_{\Phi_b}^0\left[\int\ind_{x\in\cV_0}\diff x\int \ind_{x\in\cV_0}\bE_{\Phi_{v;t}}\left[\ind_{x\in\cD(t)}\right]\diff x\right]\nnb\\
		&\ec\lambda_b\bE_{\Phi_b}^0\left[\left(\int\ind_{x\in\cV_0}\diff x\right)^2\bP(0\in\cD)\diff x\right]\nnb\\
		&=\lambda_b\bP^0(0\in\cD)\bE_{\Phi_b}^0\left[\left(\int\ind_{x\in\cV_0}\diff x\right)^2\right]\nnb\\
		&=\lambda_b\bP^0(0\in\cD)\bE_{\Phi_b}^0[(\ell_2(\cV_0))^2],\label{42}
	\end{align}
	where (a) is obtained by conditioning on the Cox point process $ \Phi_{v} $. We have (b) from the fact that $ \Phi_{v}\independent \Phi_b. $ Then, we derive (c) by using the fact that the probability of the typical point is in $ \cD(t) $ is given by $ \bP_{\Psi}^0(0\in\cD).  $ In Eq. \eqref{42}, we denote by $ \cV_0  $  the typical Voronoi cell under the Palm distribution of the base station point process. Therefore we have 
	\begin{align*}
		\bE[\ell_2(\cV_Z \cap \cD)]&=\lambda_b\bE_{\Phi_b}^0\left[(\ell_2(\cV_0))^2\right]\bP(0\in\cD),\\
		\bE[\ell_2(\cV_Z \setminus \vspace{3mm}\cD)]&=\lambda_b\bE_{\Phi_b}^0\left[(\ell_2(\cV_0))^2\right]\bP(0\notin\cD),
	\end{align*}
	respectively. Then, in \cite{hayen_quine_2002}, the second moment is given by 
	\begin{equation*}
\bE[(\ell_2(\cV_0))^2]= \frac{\nu}{\lambda_b^2}
	\end{equation*} 
	where $ J(u,v)=(\frac{1}{2}\pi + v - u)\cos(u)^2+(\frac{1}{2}\pi+u)\cos(v-u)^2+\cos(u)\cos(v-u)\sin(v) $ and 
	\begin{equation*}
		\nu = 2\pi\int_{0}^{\pi}\int_{\pi/2}^{v-\pi/2}\frac{\cos(u-v)\sin(v)\cos(u)}{J(u,v)^2}\diff u \diff v\approxeq {1.28}.
	\end{equation*}
\end{IEEEproof}

\subsection{Effective Data Rate of Typical User}
\begin{theorem}\label{Theorem:4}
	The effective data rate is given by 
	\begin{align}
		{\cT}=\dfrac{\lambda_b\int_{0}^{\infty}\bP_{\Psi}^0(\SIR>2^x-1,0\notin\cD)\diff x}{\nu\lambda_u\bP_{\Psi}^0(0\notin\cD)},\label{eq:T5}
	\end{align}
	 where the denominator is given in Theorem \ref{Theorem:1} and the numerator is given in Theorem \ref{Theorem:3}.  
\end{theorem}
\begin{IEEEproof}
The effective data rate is 
	\begin{align}
	{\cT}&= \frac{\bE_{\Psi}^0[\log_2(1+\SIR),0\notin \cD]}{\bE_{\Psi}^0[\|\Psi_{}(\cV_Y\setminus \cD(t))\|]},\nnb
	\end{align}
	where the denominator is given by 
	\begin{align}
		\bE[\|\Psi(\cV_Y\setminus \cD(t))\|]&={\lambda_u}\bE[\ell_2(\cV_Y\setminus \cD)]=\lambda_u\left(\frac{\nu}{\lambda_b}\right)\bP_{\Psi}^0(0\notin\cD).\label{33.1}
	\end{align}
	On the other hand, the numerator is 
\begin{align}
	\bE_{\Psi}^0[\log_2(1+\SIR),\text{DL}]&=\int_{0}^{\infty}\bP_{\Psi}^0(\SIR>2^x-1,0\notin\cD)\diff x.\label{33}
\end{align}
We obtain the result by combining Eq. \eqref{33.1} and \eqref{33}.
\end{IEEEproof}

	Note that the effective data rate can be interpreted as the long-term achievable rate on downlink. 

\subsection{Reliability of Sidelink in Cellular Networks}
To improve the reliability of the vehicular safety messages, one should 
 \begin{enumerate}
	\item decrease the encoding rate of the vehicular safety messages, 
	\item decrease the size of the vehicle association region, 
	\item increase the transmit power of vehicular broadcast,
	
\end{enumerate}
In practice, the encoding rate $ \epsilon $ and the size of the vehicle association region $ \rho $ might be given as the requirements for a V2X safety system. For instance, the encoding rate is determined by the size of safety messages that the ITS will adopt. Further, users are required to successfully decode such safety messages from vehicles especially from those within $ 300 $ meters. In other words, the values of $ \epsilon$ or $ \rho $ might be given as system requirements.  To improve the reliability of sidelink broadcasts, one could focus on the transmit power of the vehicular safety message $ p_v. $

Increasing the transmit power of the vehicular broadcast will improve the reliability of safety messages. Nevertheless, it is important to note that the interference generated by other vehicles will also increase. In addition, varying the transmit power of vehicles affects other network performance metrics such as the downlink coverage probability or the effective rate of the typical user. As $ p_v $ increases, the effective data rate of the typical downlink user $ \cT $ decreases due to the increased sidelink interference from vehicles. Similarly, as $ p_v $ decreases, the effective data rate will increase due to the decreased sidelink interference from vehicles. Table \ref{T:2} summarizes the tradeoff relations between the coverage probability of the sidelink and the effective rate of the downlink w.r.t. the vehicular transmit power $ p_v $. We consider $ \eta\in[0.1,1] $.
\begin{table}
	\caption{Tradeoffs in Network Performance}\label{T:2}
	\centering
	\begin{tabular}{|c|c|c|}
	\hline 
 Network parameter	& Sidelink reliability & Effective rate $ \cT $ \\ 
	\hline 
$ p_v $ increase	& increase (diminishing return) & decrease \\ 
	\hline 
$ p_v $ decrease	& decrease & increase  \\ 
	\hline 
	$ \lambda_u $ increase & unchanged & decrease \\\hline
	$ \lambda_u $ decrease & unchanged & increase \\\hline
	
\end{tabular} 
\end{table}

\begin{figure}
	\centering
	\includegraphics[width=0.8\linewidth]{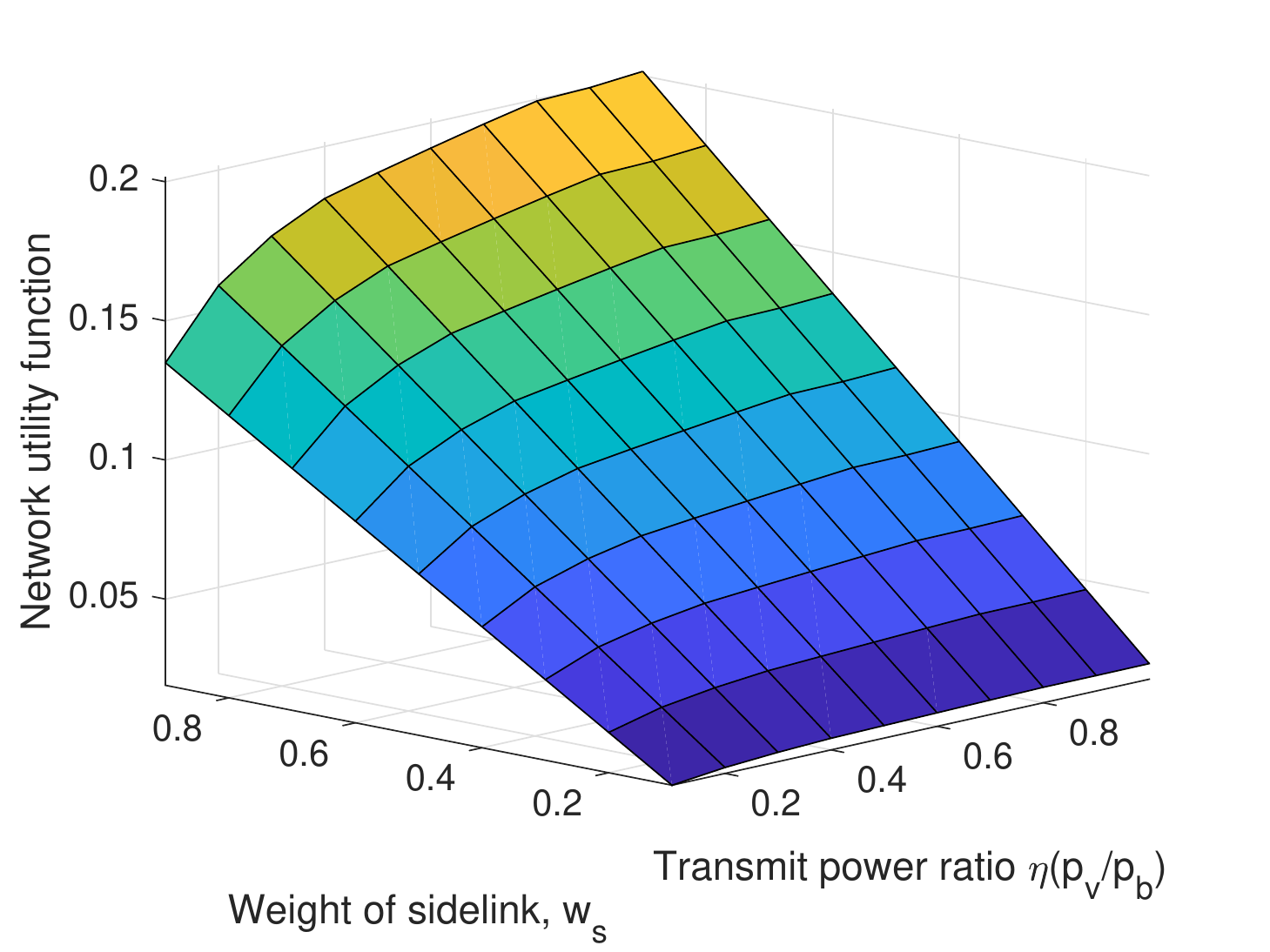}
	\caption{Network utility of Eq. \eqref{eq:networkutility} when $ \lambda_b = 5/\text{km}^2,\lambda_l=5/\text{km}, \mu=5/\text{km}, $ $ \epsilon=1, $ $ \lambda_u=200/\text{km}^2, $ and $ \alpha=3 $. }
	\label{fig:networkutility200user200metermanyweights}
\end{figure}

One way of analyzing such a tradeoff is to formulate an objective function incorporating the tradeoff. Below, we introduce the overall network {utility} given by the sum of benefits that the typical user would be offered via sidelink and downlink. For weights $ w_s, w_d >0,  $ the network utility  is the following function of $ p_v $:
\begin{align}\label{eq:networkutility}
\mathcal{U}(p_v)= w_s \underbrace{\bP(\SIR>2^\epsilon-1,\text{SL})}_{\text{SL reliability}}+ w_d\underbrace{\mathcal{T}}_{\text{DL rate}},
\end{align}
where the first term is the probability that the typical user successfully decodes the safety messages and the second term is the amount of effective data that it receives on the downlink channel. The interplay of network performances is captured by the weights, $ w_s $ and $ w_d. $
Fig. \ref{fig:networkutility200user200metermanyweights} illustrates the proposed network utility function when $ \lambda_u=200/\text{km}^2. $ To show the impact of power control in safety messaing, we consider $ \eta\in\{0.1,0.2,\ldots,1\} $ where $ \eta=1 $ indicates that $ p_v=p_b $. For each power ratio, different sidelink weights are analyzed: $ w_s\in\{0.1,0.2,\ldots,0.9\} $. In the figure, as the transmit power ratio increases, the network utility increases. The increment of network utility w.r.t. $ p_v $ is greater for a greater $ w_s $.

\begin{example}\label{corollary:2}
	By leveraging the network utility expression, we can derive the total ergodic {rate} that the typical user obtains from both sidelink and downlink. For the downlink rate, the effective rate of Eq. \eqref{eq:T5} can be used. For the sidelink rate, we use transmission capacity given in \cite{1542405}. To be specific, when the safety message is encoded with $ \epsilon  $ (bps/Hz), the probability of success is given by $ \bP(\SIR>2^{\epsilon}-1).$ Therefore, we write the sidelink rate as $ \epsilon \bP(\SIR>2^{\epsilon}-1|\text{SL}) $. Using the law of total probability, the total rate from downlink and sidelink is 
		\begin{align}
	\hat{\cT}&= \bP_{\Psi}^0(\text{SL}){\epsilon\bP_{\Psi}^0(\SIR>2^{\epsilon}-1|\text{SL})}+\bP_{\Psi}^0(\text{DL})\frac{\int_{0}^{\infty}\bP_{\Psi}^0(\SIR>2^x-1|\text{DL})}{\text{effective number of user per BS }}\nnb\\
	&= \epsilon \bP_{\Psi}^0(\SIR>2^{\epsilon}-1,0\in\cD)+ \dfrac{\lambda_b\int_{0}^{\infty}\bP_{\Psi}^0(\SIR>2^x-1,0\notin\cD)\diff x}{\nu\lambda_u\bP_{\Psi}^0(0\notin\cD)},
	\end{align}
	where $ \bP_{\Psi}^0(\SIR> 2^{x}-1,0\notin\cD) $ is in Eq. \eqref{eq:Theorem2}  and  $\bP(\SIR>2^{\epsilon}-1,0\in \cD) $ is in Eq. \eqref{eq:T3}.
\end{example}

\begin{figure}
	\centering
	\includegraphics[width=0.8\linewidth]{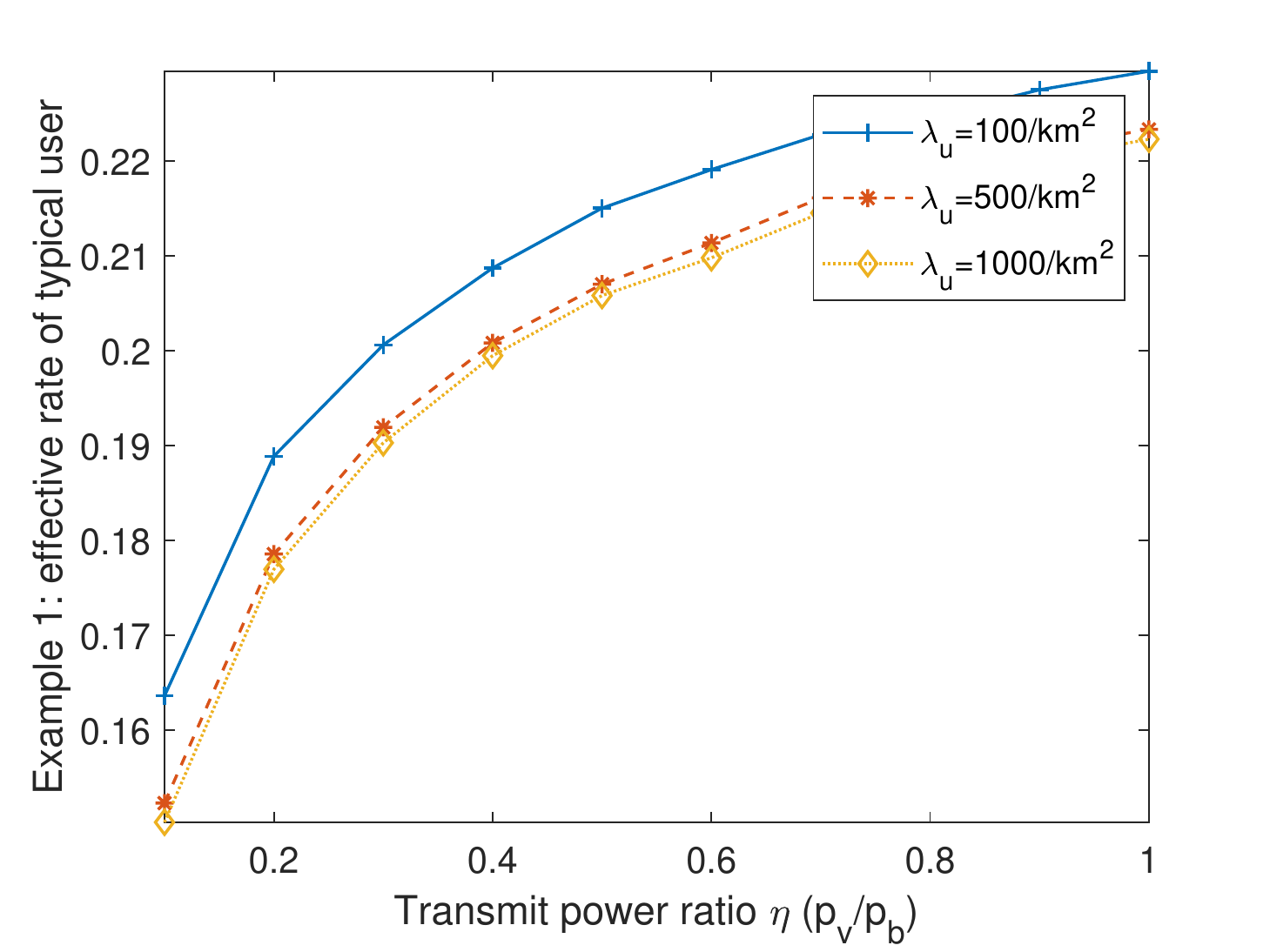}
	\caption{Total effective data rate of the typical user provided in Example \ref{corollary:2}. We consider $ \lambda_b=5/\text{km}^2, \lambda_l=5/\text{km}, \mu = 5 $/km, $ \epsilon=1 $, and $ \alpha=3 $. The mean numbers of users per base station, namely $ \lambda_u/\lambda_b $, are $ 20 $, $ 100 $, and $ 200 $, respectively. }
	\label{fig:totaleffective}
\end{figure}
Fig. \ref{fig:totaleffective} illustrates the total data rate of the typical user. The figure considers different numbers of users to analyze the impact of network congestion where the density of users is high. In the simulation setup, the effective rate increases as the power ratio increases. Furthermore, a lightly loaded network (a smaller number of users) has a better total data rate because each base station is shared by fewer users.


%

\section{Conclusion}
In this paper, we analyze the broadcast of safety messages from vehicles to users. These messages contain safety-critical information and they are assumed to be transmitted in the same spectrum as downlink communications. To capture the `compulsory' reception of vehicular safety messages from vehicles to users, we assume that users within a distance $ \rho $ are forced to decode the safety messages. To analyze the network performance, we use a Poisson line Cox point process to model vehicles and then a Boolean model on the vehicle process to represent the safety message prioritized regions. Under the proposed stationary framework, we quantify the association of users as the association probability of the typical user. We then derive the distribution of the SIR of the typical user to quantify the reliability of the vehicular safety messages and downlink messages, respectively. Finally, we derive the downlink effective rate of the typical user to quantify the impact of mandatory sidelink reception for users. To capture the tradeoff of network performances w.r.t. the transmit power of safety messages, we also analyze the network utility and the total data rate of the typical user. 
\section*{Acknowledgement}
This work is supported in part by the National Science Foundation under Grant
No. NSF-CCF-1514275 and an award from the Simons Foundation (\#197982), both
to the University of Texas at Austin.

\bibliographystyle{IEEEtran}
\bibliography{Offloading}
\end{document}